\documentclass[preprint]{JHEP3}
\JHEPspecialurl{http://jhep.sissa.it/JOURNAL/JHEP3.tar.gz}
\usepackage{epsfig,multicol,bbm}
\usepackage{citesort}
\newcommand\fverb{\setbox\fverbbox=\hbox\bgroup\verb}
\newcommand\fverbdo{\egroup\medskip\noindent%
\fbox{\unhbox\fverbbox}\ }
\newcommand\fverbit{\egroup\item[\fbox{\unhbox\fverbbox}]}
\newbox\fverbbox

\newcommand{\lsim}{\raisebox{-0.13cm}{~\shortstack{$<$ \\[-0.07cm] $\sim$}}~} 
\newcommand{\gsim}{\raisebox{-0.13cm}{~\shortstack{$>$ \\[-0.07cm] $\sim$}}~} 
\newcommand{\ra}{\rightarrow} 
\newcommand{\lra}{\longrightarrow} 
\newcommand{\tb}{\tan\beta} 
\newcommand{\beq}{\begin{eqnarray}} 
\newcommand{\eeq}{\end{eqnarray}}

\newcommand{\non}{\nonumber}

\preprint{LPT Orsay 13-26}

\title{The MSSM Higgs sector at a high $\mathbf{M_{SUSY}}$: reopening the low 
tan$\beta$ regime and heavy Higgs searches}

\author{Abdelhak Djouadi and J\'er\'emie Quevillon   \\
Laboratoire de Physique Th\'eorique, U. Paris XI and CNRS,
F--91405 Orsay, France.
}

\abstract{  One of the main implications of the LHC discovery of a Higgs boson 
with a mass $M_h \approx 126$ GeV is that the scale of supersymmetry--breaking
in the Minimal Supersymmetric Standard  Model (MSSM) might be rather high, $M_S
\gg M_Z$. In this paper, we consider the high $M_S$ regime and study the 
spectrum  of the extended Higgs sector of the  MSSM, including the LHC
constraints  on the mass and the rates of the observed light $h$ state. In
particular, we show that in a simplified model that approximates the important
radiative corrections, the unknown scale  $M_S$ (and some  other leading SUSY
parameters) can be traded against the measured value of $M_h$. One would be then
essentially left with only two free parameters to describe the Higgs  sector,
$\tb$ and the pseudoscalar Higgs mass $M_A$, even at higher orders. The main
phenomenological consequence of these high $M_S$ values is to reopen the low
$\tb$ region,  $\tb\! \lsim\! 3$--5,  which was for a long time buried under the
LEP constraint on the lightest $h$ mass when a low SUSY scale was assumed. We
show that, in this case, the heavier MSSM neutral $H/A$ and charged $H^\pm$
states can be searched for in  a variety of interesting final states such as 
decays into gauge and lighter Higgs bosons (in pairs on in mixed states) and
decays into heavy top quarks. Examples of sensitivity on the $[\tb, M_A]$ 
parameter space at the LHC  in these channels are given.}


\begin{document} 

\renewcommand{\thefootnote}{\arabic{footnote}}
\setcounter{footnote}{0}
\vspace*{-10mm}
\section{Introduction}

The discovery of a Higgs--like particle by the ATLAS and CMS collaborations 
\cite{Discovery} in July 2012 was a triumph for the Standard Model (SM) of
particle physics  as it led to a first verification of one of its cornerstones,
the Higgs sector \cite{Higgs,HHG,Review1} that realises the breaking of the
electroweak symmetry  and generates the masses of the fundamental particle
masses. This discovery had also very important  consequences  on theories beyond
the SM, among which  supersymmetric theories (SUSY) stood as the most promising
ones. This is particularly the case of  their minimal low energy realization,
the Minimal Supersymmetric Standard Model (MSSM) \cite{SUSY,pMSSM} in which the
electroweak symmetry breaking  sector is extended to contain two Higgs doublets
which,  after symmetry breaking, lead to the existence of five physical states: 
two CP--even $h$ and $H$, one CP--odd $A$ and two charged $H^\pm$ bosons
\cite{HHG,Review2}. 

The Higgs observation at the LHC with a mass of approximately 126 GeV first gave
support to the MSSM in which the lightest CP--even $h$ boson  was predicted to
have a mass  less than $\approx 130$ GeV \cite{CR-review}. An annoying  problem
is that the measured mass value is too close to the  predicted upper limit on
$M_h$ in the MSSM, suggesting that the SUSY scale is rather high, $M_S \gsim 1$
TeV; see  for  instance the discussion of  Ref.~\cite{paper1}. The fact that 
$M_S$ is large is backed up by direct SUSY particle searches, which set limits
of the order of 1 TeV for the strongly interacting superparticles
\cite{LHC-all}. In addition, with the precision measurements of its couplings to
fermions and gauge bosons, the Higgs state looked more and more SM--like, as no
significant deviations from the SM expectation  is presently observed 
\cite{LHC-all}. Although this had to be expected since, as is the case in many
extended Higgs sectors, there is a decoupling limit \cite{decoupling} in which
all the heavier Higgs particles decouple from the SM spectrum and one is left
only with the lightest $h$ state which has almost the SM properties, this is
again unfortunate. Tests of the properties of the observed Higgs state have to
be pursued with more accuracy in order to pin down small   deviations from the
SM  prediction.

An equally important way to probe the MSSM  is to search for the direct
manifestation of the heavier $H,A$  and $H^\pm$ states. These searches are
presently conducted by the ATLAS and CMS collaborations in the regime where 
$\tb$, the ratio of the vacuum expectations values of the two Higgs fields, is
very large, $\tb \equiv v_2/v_1 \gsim $5--10,  which significantly enhances the
Higgs production rates at the LHC.  The regime with low $\tb$, $\tb \lsim
3$--5,  is ignored, the main reason being that if the  SUSY scale should not
exceed $M_S \approx 3$ TeV  to have a still acceptable fine--tuning in the model
\cite{fine-tuning}, the $h$ mass is too low and does not match the observed 
value. More precisely, this $\tb$ region was  excluded  by the negative Higgs
searches that were performed at the  ancestor of the LHC, the LEP collider
\cite{LEP2}. 

In this paper, we reopen this low $\tb$ region by simply relaxing the  usual
assumption that the SUSY scale  should be in the vicinity of 1 TeV. In fact,
many scenarios with a very large scale $M_S$ have been considered in the recent
years, the most popular ones being  split--SUSY \cite{split} and   high--scale
SUSY \cite{high-scale}. In these constructions, the SUSY solution to the
hierarchy problem is abandoned and the masses of all the scalars of the theory
(and eventually also those of the spin--$\frac12$ superparticles in high--scale
SUSY) are set to very high values, $M_S \gg M_Z$. Hence, all the sfermions and
Higgs bosons are very heavy,  except for a light SM--like Higgs boson whose mass
can be as low as  $M_h \approx 120$ GeV even if $\tb$ is very close to unity. In
fact, for this purpose, the scale $M_S$ needs not to be extremely high, for
instance close to the unification scale  as in the original scenarii of
Refs.~\cite{split,high-scale}, and values of $M_S$ of order 10 to 100 TeV
would be sufficient.

In addition, one may  assume that only the sfermions are very heavy and not the
Higgs particles, as it would be the case  in non--universal Higgs models
where the soft--SUSY breaking mass parameters for the sfermion and the two
Higgs doublet fields are disconnected \cite{NUHM}. One would have then a scenario
in which the entire MSSM Higgs sector is kept at the  electroweak scale, while
the sfermions are pushed to the high scale. Such scenarios are also being
considered \cite{slim} and they might occur in many theoretical constructions. 

A first important aspect that we will address in this paper is the treatment of 
the radiative corrections in the Higgs sector and the derivation of the
superparticle and Higgs spectrum in these high scale scenarios. It is well known
that for $M_S$ values in the multi--TeV range, the MSSM spectrum cannot be
obtained in a reliable way using the usual RGE codes that incorporate the higher
order effects \cite{Suspect,RGE}: one has first to decouple properly the heavy
particles and to resum the large logarithmic contributions. Such a program has
been performed in the case where $M_A \approx M_S \gg M_Z$ and the results have
been implemented in one of the RGE codes \cite{bds}. In the absence of such a
tool for  $M_S \gg M_A \approx M_Z$ (that is under development \cite{prepa}), 
we will adopt the simple approach where the radiative corrections in the Higgs
sector are approximated by the dominant contribution in the top and stop sector,
which involves the logarithm of the scale $M_S$ and the stop mixing parameter
\cite{CR-1loop}. We will show that, in this approach, the situation simplifies
to the extent that one can simply trade the  dominant radiative correction 
against the actual value of the mass of the lighter $h$ boson that has been
measured at the LHC to be $M_h \approx 126$ GeV. An approach that is similar 
in spirit has also been advocated in Ref.~\cite{Maiani:2012ij}.

 One would then deal with a very simple post--$h$ discovery model in which, to a
very good approximation, there are only two input parameters in the Higgs
sector, $M_A$ and $\tb$  which can take any value (in particular low values $\tb
\approx 1$ and $M_A \approx 100$ GeV unless  they are excluded by the
measurements of the $h$  properties at the LHC) with the mass $M_h$ 
fixed to its measured value. If one is mainly concerned with the MSSM Higgs
sector, this allows to perform rather model--independent studies of this sector.

We should note that while the working approximation for the radiative
corrections to Higgs sector is  important  for the determination of the correct
value of $M_S$ (and eventually some other supersymmetric parameters such as the
mixing in the stop sector), it has little impact on Higgs phenomenology, i.e. on
the MSSM Higgs masses and couplings. 

The reopening of the low $\tb$  region allows then to consider  a plethora of
very interesting Higgs channels to be investigated at the LHC: heavier CP--even
$H$ decays into massive gauge  bosons $H\to WW,ZZ$ and Higgs bosons $H\to hh$,
CP--odd Higgs decays into a vector and  a Higgs boson, $A \to hZ$, CP--even and
CP--odd Higgs decays  into top quarks, $H/A \to t \bar t$, and even  charged
Higgs decay $H^\pm \to Wh$. Many search channels discussed  in the context of a
heavy SM Higgs boson or for resonances in some non--SUSY beyond the SM (new
gauge bosons or Kaluza--Klein excitations)  can be used to search for these
final states. A detailed discussion of the Higgs cross sections times decay
rates in these process is made in this paper and an estimate of the sensitivity
that could be achieved at the present $\sqrt s=8$ TeV run with the full data set
is given. These processes  allow to cover  a  large part of the parameter space
of the MSSM Higgs sector in a model--independent way, i.e. without using the
information on the scale $M_S$ and more generally on the SUSY particle spectrum
that appear in the radiative  corrections. 

The rest of the paper is organised as follows. In the next section, we discuss
the  radiative corrections in the Higgs sector when $M_h$ is used as input and
their impact on the Higgs masses and couplings. In section 3, we summarize the
various processes for  Higgs production and decay in the high and low $\tb$
regions and, in section 4, their implications for the  MSSM parameter space. In
section 5, we discuss the important  new heavy Higgs channels that can be probed
at the LHC at low $\tb$. A conclusion is given in section 6. 


\section{The Higgs sector of the MSSM in the various $\tb$ regimes}

In this section, we review the theoretical aspects of the MSSM Higgs sector with
some emphasis on the  properties of the Higgs particles in the low  $\tb$
regime, $1 \lsim \tb \lsim 3$, which contrary to the high $\tb$ regime, has not
received much attention in the literature. 


\subsection{The radiatively corrected Higgs masses} 
\label{sec:2.1}

Let us begin by recalling a few basics facts about the MSSM and its extended
Higgs sector.  In the MSSM,  two chiral superfields with respective hypercharges
$-1$ and $+1$ are needed for the cancellation of chiral anomalies  and  their
scalar components, the two doublet fields $H_1$ and $H_2$, give separately
masses to the isospin $-\frac{1}{2}$ and $+\frac{1}{2}$ fermions in a SUSY
invariant way. After spontaneous symmetry breaking,  the two doublet fields lead
to five Higgs particles: two CP--even $h,H$ bosons, a pseudoscalar $A$ boson 
and two charged $H^\pm$ bosons \cite{HHG,Review2}.  

The Higgs sector should be in principle described by the four Higgs boson 
masses  and by two mixing angle $\alpha$ and $\beta$, with $\alpha$ being the
angle which diagonalises the mass matrix of the  two CP--even neutral $h$ and
$H$ states while $\beta$ is given in terms of  the ratio of vacuum expectation
values of the two Higgs fields $H_1$ and $H_2$, $\tan\beta= v_2/v_1$. However,
by virtue of  SUSY,  only two parameters are needed to describe the  system at 
tree--level. It is common practice to chose the two basic inputs to be  the
pseudoscalar mass $M_A$, expected to lie in the range between $M_Z$  and the
SUSY breaking scale $M_S$, and the ratio $\tan\beta$, which is expected to take
values in the range \cite{tb-review}
\beq
 1 \lsim \tan\beta \lsim  \bar m_t/ \bar m_b \approx 60
\eeq 
with $\bar m_t$ and $\bar m_b$  the running top and bottom quark masses in the 
$\overline{\rm MS}$ renormalisation scheme evaluated at a scale close to the 
SUSY scale $M_S$. 

At tree--level, the CP--even $h$ boson  mass is then bound to be lighter than
the $Z$ boson, $M_h \leq {\rm min}(M_Z,M_A) |\cos2\beta| \leq M_Z$, while the
heavier $H$ and  $H^\pm$ boson have masses that are comparable to that of the
$A$ state if $M_A \gsim M_Z$.  Likewise, the mixing angle $\alpha$ can be
written in compact form in terms of $M_A$ and $\tb$. If the mass $M_A$ is large
compared to the $Z$ boson mass, the so called decoupling limit \cite{decoupling}
that we will discuss in some detail here, the lighter $h$ state
reaches its maximal mass value, $M_h \approx M_Z |\cos2\beta|$,  the heavier
CP--even   and CP--odd and the charged Higgs states become almost  degenerate in
mass, $M_H \approx  M_A  \approx M_{H^\pm}$, while the mixing angle $\alpha$
becomes close to  $\alpha \approx \frac{\pi}{2} -\beta$.    

As is well known this simple pattern is spoiled when one includes the radiative
corrections which have been shown to be extremely  important
\cite{CR-review,CR-1loop,PBMZ,CR-eff,CR-2loop,CR-FH,CR-3loop}. Once these
corrections are included, the Higgs masses (and their couplings) will, in
principle, depend on all the MSSM parameters. In the  phenomenological MSSM
(pMSSM) \cite{pMSSM}, defined by the assumptions that all the soft--SUSY
breaking parameters  are real with the matrices that eventually describe them
being diagonal  (and thus, there is no new source of CP or flavor violation) 
and by the requirement of universal parameters for the first and second
generation sfermions, the Higgs sector will   depend on, besides 
$M_A$ and $\tb$,  20 additional parameters: the higgsino
 mass parameter $\mu$; the bino, wino and gluino mass parameters
$M_1, M_2, M_3$; the first/second and third generation left-- and
right--handed sfermion mass  parameters $m_{\tilde{q}}, m_{\tilde{u}_R},
m_{\tilde{d}_R},  m_{\tilde{l}}, m_{\tilde{e}_R}$ and  $m_{\tilde{Q}},
m_{\tilde{t}_R}, m_{\tilde{b}_R},  m_{\tilde{L}}, m_{\tilde{\tau}_R}$; and
finally the (common) first/second and third
generation trilinear   $A_u, A_d, A_e$ and $A_t, A_b, A_\tau$
couplings\footnote{
The first/second generation
couplings  have no impact in general and can be ignored in practice, reducing
the effective number of free inputs of the pMSSM, from 22 to 19 parameters.}.

Fortunately, only a small subset of these parameters has a significant impact 
on the radiative corrections to the Higgs sector. At the one loop level,
the by far dominant correction to the Higgs masses is originating from top and
stop loops and grows like  the fourth power of the top quark mass,
logarithmically with the stop masses and quadratically with the stop trilinear
coupling.  The leading component of this correction reads\footnote{Note
the typographical error for this equation in Ref.~\cite{Review2} which 
translated to Refs.~\cite{paper1,paper2,paper3}.} \cite{CR-1loop}
\beq 
\epsilon = \frac{3\, \bar{m}_t^4}{2\pi^2 v^2\sin^ 2\beta} \left[ \log
\frac{M_S^2}{\bar{m}_t^2} + \frac{X_t^2}{M_S^2} \left( 1 -
\frac{X_t^2}{12\,M_S^2} \right) \right] \label{epsilon} 
\eeq 
where $\bar{m}_t$ is again the running ${\rm \overline{MS}}$ top quark mass to
account for the leading two--loop QCD and electroweak  corrections in a
renormalisation group (RG) improvement (some higher order effects can also be
included) 
\cite{CR-eff}. We  have defined  the
SUSY--breaking  scale $M_S$ to be the geometric average of the two stop masses  
$M_S = \sqrt{ m_{\tilde t_1} m_{\tilde t_2}}$; this scale is generally kept  in
the vicinity of the TeV scale to minimize the amount of fine tuning. We have
also introduced the  stop mixing parameter $X_t = A_t -\mu \cot\beta$, that we
define here in the  $\overline{\rm DR}$ scheme, which plays an
important  role and maximizes the radiative correction when   
\beq 
X_t = \sqrt 6 \, M_S \ : \ {\rm maximal~mixing~scenario} 
\label{max-mix}
\eeq
while  the radiative corrections is smallest for a vanishing $X_t$ value,   i.e.
in the no mixing scenario $X_t=0$.   An intermediate scenario is when $X_t$ is
of the same order as the SUSY scale, $X_t=M_S$, the typical mixing scenario.
These scenarios have been  often used in the past as benchmarks for MSSM Higgs
studies \cite{benchmark}  and have been  updated recently
\cite{benchmark-update}.

The $\epsilon$ approximation above allows to write the masses of CP--even Higgs 
bosons in a particularly simple form
\begin{eqnarray}
M_{h,H}^2 = \frac{1}{2} (M_A^2+ M_Z^2+\epsilon) \left[ 1 \mp 
\sqrt{1- 4 \frac{ M_Z^2 M_A^2 \cos^2 2\beta +\epsilon ( M_A^2 \sin^2\beta +
 M_Z^2 \cos^2\beta)} {(M_A^2+ M_Z^2+\epsilon)^2} } \right] \ 
\label{Mhepsilon}
\end{eqnarray}
In this approximation, the charged Higgs mass does not receive radiative 
corrections, the leading contributions being of ${\cal O} (\alpha m_t^2)$
and one can still write the tree-level relation $M_{H^\pm}=\sqrt{M_A^2+M_W^2}$.
For large values of  the pseudoscalar Higgs boson mass,  the CP--even Higgs
masses can be expanded  in powers of $M_Z^2/M_A^2$ to obtain at first order
\beq
M_h^2 & \stackrel{\small M_A \gg M_Z} \to & (M_Z^2\cos^22 \beta + \epsilon 
\sin^2\beta) \left[ 1 + \frac{ \epsilon M_Z^2 \cos^2\beta} {M_A^2 (M_Z^2 + 
\epsilon \sin^2\beta)}-\frac{M_Z^2 \sin^2\beta + \epsilon \cos^2\beta} {M_A^2}
\right] \nonumber  \\
M_H^2 & \stackrel{\small M_A \gg M_Z} \to  & M_A^2 \left[ 1 + \frac{M_Z^2 \sin^2 2\beta 
+ \epsilon \cos^2 \beta}{M_A^2}  \right] \  \ \ \ 
\eeq
and indeed, in exact decoupling $M_A/M_Z \to \infty$, one would have $M_H\! =\!
M_A \! = M_{H^+}$ for the heavier Higgs states and, for the lighter $h$ boson,
the well known relation
\beq 
M_h\! \equiv \! M_h^{\rm max}\!= \! \sqrt{M_Z^2\cos^22 
\beta \! + \!\epsilon \sin^2\beta} \label{Mhmax}
\eeq
In view of the large value $M_h \approx 126$ GeV  of the observed Higgs state at
the LHC, it is clear that some optimization of  the various  terms that enter
the mass formula eq.~(\ref{Mhmax}) with  the radiative correction
eq.~(\ref{epsilon}) is required. As was discussed in many instances including
Refs.~\cite{paper1,paper2,paper3},  one  needs:   $i)$ to be close to the
decoupling limit $M_A \gg M_Z$ and to have significant $\tb$ values that lead to
$|\cos2\beta| \to 1$ to maximize the tree--level mass  and, $ii)$  to be
in the maximal mixing scenario $X_t=\sqrt 6 M_S$ with the largest possible value
of the SUSY--breaking scale $M_S$  to maximize the radiative corrections. As the other
SUSY--breaking parameters do not affect  significantly the $M_h^{\rm max}$ value,
one can  fix them to some value. For instance, one can make the choice
\cite{benchmark-update}
\begin{eqnarray}
{M_h}^{\rm max}_{\rm bench} 
 : \begin{array}{c} 
M_2  \simeq 2 M_1 = |\mu|=\frac15 M_S, \, M_3= m_{\tilde q_i}=\frac 13
m_{\tilde \ell_i} = 1.5 M_S, \, A_i =0, \\
m_{\tilde b_R} = \frac13 m_{\tilde \tau_i} = M_S \, , \ A_b=A_\tau=A_t 
\end{array}
\label{pbenchmark}
\end{eqnarray}
where $m_{\tilde q_i}$ and $m_{\tilde \ell_i}$ are the common first/second 
sfermion SUSY--breaking masses and $A_i$ their  trilinear  couplings.
Alternatively, one can perform a scan of these parameters in a reasonable range
which should change the resulting value of $M_h^{\rm max}$ in the 
$\overline{\rm DR}$ scheme only by a few GeV in general. 

In the case of a not too large SUSY scale, $M_S \lsim 3$ TeV, the numerical 
analyses of the MSSM Higgs sector can be performed with RGE programs 
\cite{Suspect,RGE} such as {\tt Suspect} which include the most relevant higher
order  radiative corrections in the calculation of  the Higgs and superparticle
masses  (and their couplings). In  particular, for the Higgs sector,  the full
set of one--loop radiative  corrections which  include also the sbottom and stau
loop corrections that are important at high $\tb$ values \cite{PBMZ} and  the
dominant two--loop QCD and electroweak corrections   \cite{CR-2loop} are
incorporated  in the $\overline{\rm DR}$ scheme; the dominant three--loop
corrections are also known \cite{CR-3loop} but they are quite small and they can
be neglected.

One should compare the results with those obtained with the program {\tt
FeynHiggs} \cite{Feynhiggs} which incorporates the radiative corrections at the
same level of accuracy but in the on--shell  renormalisation scheme 
\cite{CR-FH}. In most cases, one obtains comparable results but in some scenarios,
the difference in the values of $M_h$ can be as large as 3 GeV. We will thus
assume, as in Ref.~\cite{benchmark-update}, that there is a $\Delta M_h \approx
3$ GeV uncertainty on the determination of the $h$ mass in the MSSM and that the
value $M_h = 126$ GeV of the particle observed at the LHC corresponds to a
calculated mass within the pMSSM of 
\beq
123~{\rm GeV} \leq M_h \leq 129~{\rm GeV} \label{Mherror}
\eeq 
This uncertainty includes the parametric uncertainties of the SM inputs, in 
particular the $\overline{\rm MS}$ $b$--quark mass and the top quark pole mass 
$\overline{m}_b (m_b)\!= \!4.7~{\rm GeV}$  and $m_t^{\rm pole}\! = \!173.2 \pm
1~{\rm GeV}$ \cite{PDG}. In the latter case, it is assumed that the top quark
mass measured at the Tevatron, with the uncertainty of 1 GeV,  is indeed the
pole mass. If the top mass is instead extracted from the top pair production
cross  section, which provides a theoretically less ambiguous  determination of 
$m_t^{\rm pole}$, the uncertainty would be of order 3 GeV \cite{alekhin-fate}.
Including also the experimental error in the $M_h$ measurement by   ATLAS and
CMS, $M_h=125.7\pm 0.4$ GeV,  the possible  calculated mass
value of  the $h$ boson  in the MSSM can be extended to the much wider  and
admittedly rather  conservative range $120\;{\rm GeV}\! \leq \! M_h \! \leq \! 
132\;{\rm GeV}$. 

\subsection{The low $\tb$ regime}

The previous discussion assumed a not too high SUSY--breaking scale, $M_S \lsim
3$ TeV, in order not to have a too large fine-tuning in the model. However, in
many scenarios, values of $M_S$ in the 10 TeV range  and even beyond have been 
considered, with a most popular one being the split--SUSY scenario
\cite{split,strumia}.  Indeed, as the criterion to quantify the acceptable
amount of tuning is rather subjective, one could well have a very large value of
$M_S$ which implies  that no sfermion is accessible at the LHC or at any
foreseen collider, with the immediate advantage of solving the flavor and CP
problems in the MSSM by simply decoupling these states. The  mass parameters for
the spin--$\frac12$ particles,  the gauginos and the  higgsinos, can be kept
close to the electroweak  scale,  allowing for a solution to the dark matter 
problem and a successful gauge coupling unification, the two other SUSY
virtues.  The SUSY solutions to these two remaining problems are abandoned if
one takes the very extreme attitude of assuming that the gauginos  and higgsinos
are also very heavy, with a mass close to the scale $M_S$, as  is the case of the
so--called high--scale SUSY models \cite{high-scale,strumia}.  

In all these these SUSY scenarios, there is still a light particle, the $h$
boson, which can have a mass close to 126 GeV for a given choice of parameters
such as $M_S$ and $\tb$; see for instance Refs.~\cite{paper1,strumia}. The other
Higgs particles are much heavier as the pseudoscalar Higgs mass is very often
related to the mass scale of  the scalar fermions of the theory,  $M_A \approx
M_S$. However, this needs  not to be the case in general,  in particular for
$M_S$ values not orders of magnitude larger than 1 TeV. Even, in constrained
minimal Supergravity--like scenarios, one can assume that the soft
SUSY--breaking scalar mass terms are  different for the sfermions and for the
two Higgs doublets, the so--called non--universal Higgs mass models \cite{NUHM}
in which the mass $M_A$ is decoupled from $M_S$. Scenarios with very large
values of $M_S$ and values of $M_A$ close to the weak scale have been advocated
in the literature \cite{slim}, while models in which one of the soft
SUSY--breaking Higgs mass parameters, in general $M_{H_1}$, is at the weak scale
while $M_S$ is large are popular; examples are the  focus point scenario
\cite{focus} and  the possibility  also occurs in M/string theory inspired
scenarios \cite{mtheory}.   

Hence, if one is primarily concerned with the MSSM Higgs sector, one may be
rather conservative and assume any value for the pseudoscalar Higgs mass $M_A$ 
irrespective of the SUSY scale $M_S$. This is the quite ``model--independent"
approach that we advocate and will follow in this paper: we take $M_A$ as a free
parameter of the pMSSM, with values ranging from slightly above 100 GeV up to
order $M_S$, but make no restriction on the SUSY scale  which can be set to any
value.

Nevertheless, in scenarios with $M_S \gg 1$ TeV, the Higgs and SUSY mass
spectrum cannot be calculated reliably using standard  RGE programs  as one  has
to properly decouple the heavy states from the low-energy theory and resum the
large logarithmic corrections. A comprehensive study of the split SUSY spectrum
has  been performed in Ref.~\cite{bds} and the various features  implemented in
an adapted version of the  code {\tt SuSpect}. However, this version does not
include the possibility $M_S \gg M_A \gsim M_Z$ that is of interest for us
here.  A comprehensive and accurate description of the high $M_S$ scenario in
the MSSM in the light of the $h$ discovery, including the possibility of a Higgs
sector at the weak scale,  is under way \cite{prepa}. In the meantime,  we will
use the  $\epsilon$ approximation of eq.~(\ref{epsilon}) to describe the
radiative corrections in our high $M_S$ scenario which should be a good
approximation for our purpose. In particular, for  $M_A \gg M_Z$, we have
verified that our results  are in a relatively good agreement with those derived
in the more refined approach of Ref.~\cite{bds}.

\begin{figure}[!h]
\begin{center}
\epsfig{file=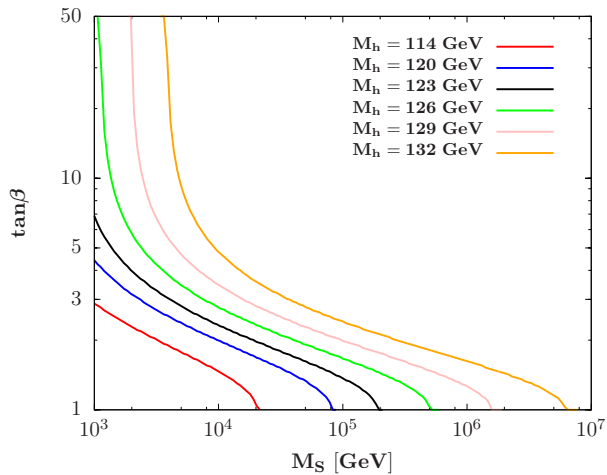,height=6.2cm} 
\end{center}
\vspace*{-5mm}
\caption[]{Contours for fixed values $M_h=120, 123, 126,129$ and 132 GeV
in the $[\tb, M_S]$ plane in the decoupling limit  $M_A \gg M_Z$; the ``LEP2 
contour" for $M_h=114$ GeV is also shown.}
\label{Fig:mass}
\end{figure}

Let us now discuss the magnitude of the  SUSY scale that is needed to make small
$\tb$ values viable. We make use of the program {\tt  Suspect} in which the
possibility $M_S \gg 1$ TeV is implemented \cite{bds} and which includes the
full set of radiative corrections  (here we assume  the maximal mixing $X_t
=\sqrt 6 M_S$ scenario and we take 1 TeV for the gaugino and higgsino masses).
In  Fig.~\ref{Fig:mass}, displayed are the contours in the plane $[\tb, M_S]$
for  fixed mass values $M_h=120, 123, 126,129$ and 132 GeV of the observed Higgs
state (these include the 3 GeV theoretical uncertainty and also a 3 GeV
uncertainty on the top quark mass).

From the figure, one concludes that values of $\tb$ close to unity are possible
and allow for an acceptable $M_h$ value provided the scale $M_S$ is large
enough. For instance,  while one can accommodate a scale $M_S \approx 1$ TeV  
with $\tb \approx 5$, a large scale $M_S \approx 20$ TeV is required to reach 
$\tb \approx 2$; to reach the limit $\tb=1$, an order of magnitude increase of
$M_S$ will be needed. Outside the decoupling regime, the  obtained $M_S$ for a
given $M_h$ value will be of course larger. For completeness, we also show the
contour for the mass value $M_h=114$ GeV, the 95\% confidence level limit
obtained at LEP2 on a SM--like Higgs boson; it illustrates the fact that values
down to $\tb \approx 1$ are still allowed by this bound provided that $M_S \gsim
10$ TeV. The implications of this feature will be discussed later.

In the rest of this paper, we will thus consider situations with the MSSM Higgs
sector at the weak scale and the only requirement that we impose is that it
should be compatible with the LHC data and, in particular, with the mass and
production rates of the Higgs boson that  has been observed.   The requirement
that $M_h\! \approx\! 126 $ GeV, within the theoretical and experimental
uncertainties, will be turned into a requirement on the  parameters that enter
the radiative corrections and, hence, on the  scale $M_S$ and the mixing
parameter $X_t$, for given values of the two basics inputs $M_A$ and $\tb$.  

\subsection{The Higgs couplings and the approach to the decoupling limit}

Let us now turn to the important issue of the Higgs couplings  to fermions and
gauge bosons. These couplings strongly depend on $\tb$ as well as on the angle
$\alpha$ (and hence on $M_A$);  normalized to the SM Higgs couplings, they are
given in Table \ref{Hcoup}.  The  $A$ boson  has no tree level couplings to
gauge bosons as a result of CP--invariance, and its couplings to down--type and
up--type fermions are, respectively, proportional and inversely proportional to
$\tb$. This is also the case for the couplings of the charged Higgs boson to
fermions, which are admixtures of $\bar m_b \tb$ and $\bar m_t \cot \beta$ terms and
depend only on $\tb$. For the CP--even Higgs bosons $h$ and $H$, the couplings
to  fermions are ratios of sines and cosines of the angles $\alpha$ and $\beta$;
the couplings to down (up) type  are enhanced (suppressed) compared to the SM
Higgs couplings for $\tb >1$. The two states share the SM Higgs couplings to
vector bosons as they are suppressed by $\sin(\beta-\alpha)$ and
$\cos(\beta-\alpha)$, respectively for $h$ and $H$. We note that there
are also couplings between a gauge and two Higgs bosons which  in the case of
the CP--even states are complementary to those to two gauge bosons $g_{h A Z}
\propto  g_{ h H^+ W^-} \propto g_{HVV}$ and vice  versa for $h\leftrightarrow
H$; the coupling  $g_{A H^+ W^-}$ has full strength.

\begin{table}[htbp]
\renewcommand{\arraystretch}{1.3}
\begin{center}
\begin{tabular}{|c|c|c|c|c|} \hline
$\  \Phi  \ $ & $ g_{\Phi \bar{u}u} $    &   $ g_{\Phi \bar{d} d} $  &
$g_{ \Phi VV} $  & $g_{ \Phi A Z} / g_{ \Phi H^+ W^-} $ \\ \hline
$h$ & $ \cos\alpha/ \sin\beta $  &  
$-\sin\alpha/ \cos\beta  $  & $ \sin(\beta-\alpha) $ & $ \propto \cos(\beta-\alpha) $ \\ \hline
$H$ & $\sin\alpha/ \sin\beta$  & $\cos\alpha / \cos\beta$ & 
$\cos(\beta-\alpha)$ &  $ \propto \sin(\beta-\alpha) $ \\ \hline
$A$  & \ $\; {\rm cot}\beta \; $\ & \ $ \; \tb \; $ \   & \ $ \; 0 \; $ & 
$ \propto 0/1$ \\  \hline
\end{tabular}
\end{center}
\vspace*{-1mm}
\caption{The couplings of the neutral MSSM Higgs bosons, collectively 
denoted by $\Phi$,  to fermions and gauge bosons when normalized to the SM 
Higgs boson couplings.}
\vspace*{-1mm}
\label{Hcoup}
\end{table}

These couplings are renormalized essentially by the same radiative corrections 
that affect the CP--even neutral Higgs masses. In the $\epsilon$ approximation 
discussed above, the one--loop radiatively corrected mixing angle $\bar \alpha$ 
will indeed read 
\beq
\tan 2 \bar{\alpha} = \tan 2\beta \,  \frac{M_A^2 + M_Z^2} {M_A^2 - 
M_Z^2 + \epsilon /\cos 2 \beta } \
\label{alphaCR}
\eeq
This  leads to corrected reduced $h,H$ couplings to gauge bosons that are
simply  $g_{hVV}= \sin (\beta- \bar \alpha)$ and $g_{HVV}= \cos (\beta- \bar 
\alpha)$ and similarly for the couplings to fermions.

The decoupling limit is controlled by the $VV$ coupling of the heavier  CP--even
Higgs boson, $g_{HVV} = \cos(\beta -\bar \alpha)$, which vanishes in this case,
while  the $hVV$ coupling $g_{hVV}^2=1- g_{HVV}^2= \sin^2(\beta-\bar \alpha)$
becomes SM--like. Performing again an expansion in terms of the pseudoscalar
Higgs mass, one obtains in the approach to the decoupling limit\footnote{We 
thank Nazila Mahmoudi for discussions and help concerning  these limits.} 
\beq
g_{HVV}   \stackrel{\small M_A \gg M_Z} \lra \ \chi \equiv 
\frac12 \frac{M_Z^2}{M_A^2} \; \sin4 \beta 
- \frac12 \frac{\epsilon}{M_A^2} \;  \sin 2 \beta 
\quad    \ra 0 \label{gHVVdecoup}
\eeq
where, in the intermediate step,  the first term is due to the tree--level 
contribution and the second one to the one--loop contribution $\epsilon$.
Concentrating  first on the tree--level part, one realises that for large values
of $\tb$ and also for values very close to unity, the decoupling limit is
reached more quickly.  Indeed the expansion parameter involves also the factor
$\sin 4\beta$ which  becomes in these two limiting  cases 
\beq
\sin 4\beta = \frac{4 \tb (1- \tan^2\beta)}{(1+ \tan^2\beta)^{2} } \to  
\bigg\{
\begin{array}{l}
- 4/\tb~~      {\rm for}~\tb \gg 1 \\
1- \tan^2\beta~   {\rm for}~\tb \; \sim \, 1 
\end{array}
\ \to 0 \ 
\eeq
Hence, in both the $\tb \gg 1$ and $\tb \sim 1$ cases, the $g_{HVV}$ coupling 
that controls the decoupling limit $M_Z^2/M_A^2 \to 0$, is doubly suppressed.
The radiatively generated component, if one recalls that  the one--loop
correction in eq.~(\ref{epsilon}) involves a $1/\sin^2 \beta$ term  which makes
it behave as $- \epsilon/M_A^2 \times \cot \beta$, also vanishes at high $\tb$
values. This leads to the well known fact that the decoupling limit $g_{HVV} \to
0$  is reached very quickly in this case, in fact as soon  as $M_A \gsim
M_h^{\rm max}$.  Instead, for $\tb \approx 1$, this radiatively generated
component is maximal.  However, when both components are included, the departure
from the  decoupling limit in the coupling $g_{HVV}$ for a fixed $M_A$ value
occurs when $\sin4\beta \approx -1$, which corresponds to $\beta = 3\pi/8$ and
hence  to the value $ \tb \approx 2.4$.

Similarly to the $HVV$ case, one can write the couplings of the CP--even
Higgs states to isospin $\frac{1}{2}$ and $-\frac{1} {2}$ fermions in the
approach to the decoupling limit $M_Z^2/M_A^2 \ll 1$ as 
\beq
\label{gHff:decoup}
\begin{array}{llll}
g_{huu} & \stackrel{\small M_A \gg M_Z} \lra \  & 1 + \chi \; \cot \beta  
& \to 1  \nonumber \\ 
g_{hdd} & \stackrel{\small M_A \gg M_Z} \lra \  &1 - \chi \; \tb  
\  & \to 1 \non \\ 
g_{Huu} & \stackrel{\small M_A \gg M_Z} \lra \ & -\cot\beta + \chi
& \to  -\cot \beta \non \\ 
g_{Hdd} & \stackrel{\small M_A \gg M_Z} \lra \ & + \tb + \chi
& \to + \tb
\end{array}
\eeq 
with the expansion parameter $\chi \propto 1/M_A^2$ is the same as the one given
in eq.~(\ref{gHVVdecoup}).  In the $M_A \gg M_Z$ regime, the couplings of the
$h$ boson approach those of the SM  Higgs boson,  $g_{huu} \approx g_{hdd}
\approx1$, while the couplings of the  $H$ boson reduce, up to a  sign, to those
of the pseudoscalar Higgs boson, $g_{Huu} \approx g_{Auu}  = \cot \beta$ and
$g_{Hdd} \approx g_{Add} = \tb$. Again, as a result of the presence of the same
combination of  $M_Z^2\sin4\beta$ and $\epsilon  \sin 2\beta$  factors in
the expansion term  $\chi$ of all couplings, the limiting values are reached more
quickly at large values of $\tb$ but the departure from  these values is slower
at low $\tb$. 

\begin{figure}[!h]
\begin{center}
\epsfig{file=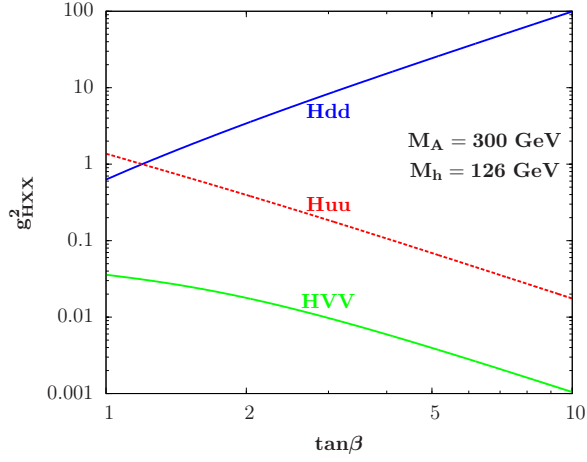,height=6cm} 
\end{center}
\vspace*{-5mm}
\caption[]{The squared couplings of the  heavier CP--even $H$ 
state to gauge bosons and fermions as a function of $\tb$ for 
$M_A=300$ GeV. The SUSY scale is chosen so that $M_h=126$ GeV.}
\label{Fig:coupling}
\vspace*{-1mm}
\end{figure}

In Fig.~\ref{Fig:coupling}, we display the square of the $H$ couplings to  gauge
bosons and fermions as a function of $\tb$ for $M_A =300$ GeV. Again the maximal
mixing scenario is  assumed and $M_S$ is chosen in such way that  for any $\tb$
value, one has $M_h=126$ GeV. At such $A$ masses,  the couplings of the lighter
$h$ boson to all particles deviate little from unity even for small $\tb$ values
and in this case too  one can consider that  we are already  in the decoupling
regime. Nevertheless,  the coupling of the heavier $H$ boson to $VV$ states is
still non--zero, in particular at low $\tb$. The $H$ coupling to $t\bar t$ 
pairs states (as well as the $A$ coupling) is significant at low $\tb$ values,
$g_{Htt}^2 \gsim 0.1$ for $\tb \lsim 3$. It even becomes larger (and the $Hbb$ coupling smaller)
 than unity for $\tb \lsim 1.2$. 
 
This demonstrates that the heavier $H/A/H^\pm$ bosons can have sizable
couplings to top quarks (and to massive gauge bosons for $H$ outside the
decoupling regime)  if $\tb$ values as low as $\sim 3$ are allowed. In
fact, the $H/A/H^\pm$ couplings to top quarks $\propto \cot\beta$ are larger
that the couplings to bottom quarks $\propto \tb$ for values $\tb \approx \sqrt{
\bar m_t / \bar m_b} \lsim 7$ and this value should be considered as the
boundary between the high and low $\tb$ regimes. With more refinement, one can 
consider three $\tb$ regimes: the high regime with $\tb \gsim 10$, the
intermediate regime with $5 \lsim \tb \lsim 10$ and the low regime with $\tb
\lsim 5$. 

\vspace*{-1mm}

There are two important remarks which should be made before closing this
section.  The first one is that besides the $\epsilon$ correction,  there are 
additional one--loop vertex corrections which modify the tree--level 
Higgs--fermion couplings \cite{CR-deltab}. In the case of $b$--quarks
in the high (and eventually intermediate) $\tb$ regime,
they can be  very large in the $b$--quark case as they grow as $\bar m_b 
\tan\beta$. The dominant component comes from the SUSY--QCD corrections 
with sbottom--gluino loops that can be
approximated by
\beq 
\Delta_b \simeq 
 \frac{2\alpha_s}{3\pi} \mu m_{\tilde{g}} \tb /{\rm max}(m_{\tilde{g}}^2, 
m_{\tilde{b}_1}^2,m_{\tilde{b}_2}^2) \label{deltab}
\eeq 
In the decoupling limit $M_A \gg M_Z$, the reduced $b\bar b$  
couplings of the $H,A$ states read 
\beq
g_{Hbb} \approx g_{Abb} \approx  \tb \bigg[1-\frac{\Delta_b} {1+\Delta_b} 
\bigg] 
\label{deltabH}
\eeq
In the case of the lighter $h$ boson, the $hbb$  couplings stay SM--like in 
this limit in principle, but slightly outside the decoupling limit, there is
a combination of SUSY parameters which realises the so--called  ``vanishing
coupling" regime \cite{benchmark} in which $\bar \alpha \to 0$ and hence 
$g_{hbb} \ll 1$.  

The second remark concerns the trilinear $Hhh$ coupling which will be needed
in our analysis. In units of $M_Z^2/v$, this coupling is given 
at tree--level by \cite{HHG}
\beq
\lambda_{Hhh} \approx 2\sin2 \alpha \sin (\beta+\alpha) -\cos 2\alpha \cos(\beta
+ \alpha) 
\eeq

Again, to include the radiative corrections in the $\epsilon$ approximation, 
one needs to perform the change $\alpha \to \bar \alpha$; however, in this case,
there are also direct vertex corrections but they can be still described by the
$\epsilon$ parameter. One obtains in this approach \cite{DKMZ}
\beq
\lambda_{Hhh} \stackrel{\small M_A \gg M_Z} \lra - \frac{3}{M_Z^2}
\bigg[ \sqrt{(M_h^2 -
\epsilon \sin^2\beta)(M_Z^2 -  M_h^2  + \epsilon \sin^2\beta )}
+  \epsilon \sin \beta  \cos \beta \bigg] \label{Hhhcp}
\eeq
At high--$\tb$, the trilinear coupling vanishes $\lambda_{Hhh} \to 0$ while
for small and intermediate $\tb$ values it stays quite substantial as a
 result of the large $\epsilon$ corrections. 

\section{Higgs decays and production at the LHC }

\subsection{The high and intermediate $\tb$ regimes}

The production and decay pattern of the MSSM Higgs bosons crucially depend on
$\tb$. In the LHC run up to now, i.e. with center of mass energies
up to $\sqrt s=8$ TeV,   only relatively large  $\tb$ values, $\tb \gsim 5$--10
which correspond to the high and intermediate regimes, are probed in the search
of the neutral $H/A$ and the charged $H^\pm$ bosons. In the high $\tb$ regime, 
the couplings of these non--SM like Higgs bosons to $b$ quarks and to $\tau$
leptons are so strongly enhanced,  and the couplings  to top quarks and massive
gauge bosons so suppressed, that the  pattern becomes rather  simple.

A first simplifying feature is that the decoupling regime  in which the lighter
$h$ boson attains its maximal mass  $M_h^{\rm max}$ value for a given SUSY
parameter set\footnote{The present discussion  holds in the case where the $h$
boson is the SM--like state which implies $M_A \gsim M_h^{\rm max}$. At low 
$M_A$ values, the role of the CP--even $h$ and $H$ states are reversed:  it is
$H$  which is the SM--like particle $H\equiv H_{\rm SM}$  and $h$ would
correspond to the pseudoscalar--like Higgs particle. However,  the possibility
that the $H$ state is the observed particle at the LHC is ruled out by present
data \cite{paper3}. A special case would be  $M_A \approx M_h^{\rm max}$, which
is called the intense coupling regime  in Ref.~\cite{intense} and which leads to
mass degenerate $h,H,A$ states with comparable couplings to fermions; as the $h$
and $H$ states are close in mass, one has the same phenomenology as in the
decoupling limit where $H$ has the same properties as $A$  \cite{Baglio2}. 
Again, this scenario is strongly disfavored by present data \cite{paper3}.} and
has SM--couplings already  at $M_A \gsim M_h^{\rm max}$ for $\tb
\gsim 10$. In this case,  the heavier CP--even  $H$ boson has approximately the
same mass as the $A$ boson and its interactions are similar. Hence, the spectrum
will consist of a SM--like Higgs $h \equiv H_{\rm SM}$  and two pseudoscalar
(like) Higgs particles, $\Phi=H/A$. The $H^\pm$ boson will also be approximately
degenerate in mass with the $\Phi$ states and the intensity of its couplings to
fermions will be similar.  

An immediate consequence will be that the $h$ boson will precisely  decay into
the variety of final states and will be produced in the various channels that
are present in the SM. These decay and production processes have been studied in
detail at various places, see  Ref.~\cite{Review1} for a detailed review and
Refs.~\cite{Baglio2,LHCWG} for updates.  We will  discuss the implications  of
these channels for the properties of the state observed at the LHC in the
next section.  

In the case of the heavier neutral $\Phi=H/A$ bosons, the decay pattern is very
simple: the $t\bar t$ channel and all other decay modes are suppressed to a
level where their branching ratios are negligible and the $\Phi$ states   decay 
almost exclusively into  $\tau^+\tau^-$ and  $b\bar b$ pairs,  with branching
ratios of BR$(\Phi \to \tau^+ \tau^-) \approx  m_\tau^2/[3 \overline{m}_b^2
(M_\Phi)+m_\tau^2]\approx 10\%$ and BR$(\Phi \to b \bar b) \approx 90\%$. The
charged Higgs particles decay into $H^\pm \to \tau \nu_{\tau}$ final states with
a branching fraction of almost 100\% for $H^\pm$ masses  below the $tb$
threshold, $M_{H^\pm} \lsim m_t-m_b$, and a branching ratio of only $\approx
10\%$ for masses above this threshold.  The by far dominant channel in the
latter case is $H^\pm  \to t b$ which occurs with a $\approx 90\%$ probability
for the  same reason as above.

Concerning Higgs production in the high $\tb$ regime,  the enhancement of the
$b$--quark  couplings  makes that only processes involving this quark are
important for the $\Phi=H/A$ states.  In the dominant gluon fusion production
channel, $gg\!  \to\!  \Phi$,  one  should take into account  the $b$--quark
loop  which provides the largest contribution  (in contrast to the SM where the
top quark contribution largely dominates)  and in associated Higgs production
with heavy quarks, $b\bar{b}$ final states and hence the processes $gg/q\bar q
\to b\bar b+\Phi$, must be considered. The latter processes are equivalent to
the $b \bar b \to \Phi$ channels when no--additional $b$--quark in the final
state is present, if one considers the $b$--quark as a massless parton and  uses
heavy quark distribution functions in a five active flavor scheme \cite{scott}. 

Hence, except for the $gg \to \Phi$ and $b\bar b \to \Phi$ fusion processes, all
the other production channels are irrelevant in the high $\tb$ regime, in
particular the vector boson fusion and the Higgs--strahlung channels, that are
absent for $A$ and strongly suppressed for $H$. In both cases, as $M_\Phi \gg
m_b$, chiral symmetry holds and the cross sections are approximately the same
for the CP--even $H$ and CP--odd $A$ bosons. The cross section for $gg\to \Phi$
is known up to next--to--leading order in QCD \cite{ggH-NLO} and
can be calculated using the program {\tt HIGLU} \cite{HIGLU,Michael-web}. The
$bb\to \Phi$ rate is instead known up to NNLO in QCD \cite{bbH-NLO,bbH-NNLO} and its
evaluation can be made using the programs {\tt bb@nnlo} or {\tt SUSHI}
\cite{SUSHI}. Note that for associated $H/A$ production with two tagged
$b$--quarks in the final states that can be used, one should instead consider
the process $gg/q\bar q \to bb+\Phi$ which is known  up to NLO QCD
\cite{ggbbH-NLO}; they leading order cross section can be obtained using the
program {\tt QQH} \cite{Michael-web}.

The most powerful search channel for the heavier MSSM Higgs particles at the LHC
is by far the process 
\beq
pp\to gg+b \bar b \to \Phi \to \tau^+ \tau^-
\eeq
 The precise
values of the cross section times branching fraction for this process at the LHC
have been recently updated in Refs.~\cite{Baglio2,LHCWG} and an assessment of 
the associated theoretical uncertainties has been made. It turns out that these
uncertainties are not that small. They consist mainly of the scale uncertainties due
to the missing higher orders in perturbation theory and of the combined 
uncertainty from the parton distribution functions  and the strong coupling
constant $\alpha_s$. When combined, they lead to a total theoretical
uncertainty  of 20--30\% in both the $gg\to \Phi $ and $b\bar b \to \Phi$
channels\footnote{It was advocated in Ref.~\cite{Baglio2} that there are two
additional sources of uncertainties related to the $b$--quark mass which should
be considered: the one in the $gg\to \Phi$ process due to the choice of the
renormalization scheme for $m_b$ and  the parametric uncertainty. These could
significantly increase the total uncertainty. We will however, ignore this
complication and retain the ``official" estimate of the error given in
Ref.~\cite{LHCWG}.}. We  will assume here for the combined $gg+b\bar b \to
\Phi$ channel  a theoretical uncertainty of  
\beq   
\Delta^{\rm TH} \sigma (pp\!\to\! \Phi) \times {\rm BR} (\Phi\! \to\! \tau  
\tau)  = \pm 25\%  
\eeq  
in the entire $M_\Phi$ range probed at the LHC and for both $\sqrt s=8$ and 14
TeV.

Besides the QCD uncertainty, three other features could alter the rate $\sigma
(pp\!\to\! \Phi \! \to\! \tau \tau)$ in the MSSM and they are related to the
impact of the SUSY particle contributions. We briefly  summarise them below and
some discussions are also given in  Refs.~\cite{paper3,ABM}.

While the CP--odd $A$ state does not couple to identical squarks as a result of
CP--invariance, there is a $H \tilde q_i \tilde q_i$ coupling in the case of the
$H$ state which allows squarks, and mainly  top and bottom squarks, to
contribute to the $gg \to H$ amplitude at leading order  (there are NLO
contributions \cite{ggH-squarks} for both the $Hgg$ and  $Agg$ amplitudes via
gluino exchange  but they should be smaller). However, as squarks do not couple
to the Higgs bosons proportionally to their masses, these contributions are
damped by  powers of $\tilde m^2_Q$ for $M_H \lsim 2m^2_Q$ and,  at high $\tb$.
the $b$--loop contribution stays largely dominant. These SUSY contributions  are
thus expected to be small and can be neglected in most cases.

 A more important effect of the SUSY sector is due to the one--loop vertex 
correction  to the $\Phi b\bar b$ couplings, $\Delta_b$ of 
eqs.~(\ref{deltab}--\ref{deltabH}), which can be large  in the high  $\tan\beta$
regime as discussed  previously.  However, in the case
of the full process $p p\! \to\! \Phi\! \to\! \tau^+ \tau^-$, this correction 
appears in both the cross section, $\sigma(\Phi) \propto (1+ \Delta_b)^{-2}$, 
and in the branching fraction, BR$(\tau\tau)= \Gamma(\Phi \to
\tau\tau)/[(1+ \Delta_b)^{-2}\Gamma(\Phi \to b\bar b)+ \Gamma(\Phi \to \tau\tau)]$, which involves
the $\Delta_b$ correction  above  in the denominator. Hence, in the cross
section times branching  ratio, the  $\Delta_b$ corrections largely cancels out
and for BR$(\tau \tau)\approx 10\%$, one obtains  
\beq  \sigma (gg+b\bar b \to \Phi) \times {\rm BR}(\Phi \to \tau\tau)   
\approx  \sigma \times {\rm BR} \times (1- \frac15 \Delta_b)  
\eeq  

Hence, one needs a very large $\Delta_b$ term (which, one should recall, is a 
radiative correction and should be small, for a recent discussion, see for instance 
Ref.~\cite{Liu:2012qu}), of order unity or more, in order to alter significantly
the $pp \to \Phi \to \tau\tau$ rate\footnote{In any case, if one insists to take
this $\Delta_b$ correction into account in the constraint on the $[\tb,M_A]$
plane  that is obtained from the $pp\to\Phi\to \tau\tau$ rate, one could simply
replace $\tb$  by $\tb /(1+\Delta_b/10)$. A contribution $\Delta_b \approx 1$ 
will change the limit on $\tb$ by only 10\%, i.e. less than the QCD
uncertainty.}. 

Finally, there is the possibility that there are light SUSY particles with 
masses  $\widetilde m \lsim \frac12  M_{\Phi}$ which lead to the opening of 
SUSY decay channels for the $H/A$ states  that might reduce the $\Phi \to \tau
\tau $ branching fraction. For $M_\Phi \lsim 1$  TeV, the only possibilities for
these superparticles seem to be light neutralinos and charginos ($\chi$) and
light sleptons $(\widetilde \ell)$. These decays have been reviewed in
Ref.~\cite{Review2} and they have been found to be in general disfavored in the
high $\tb$ regime as the $\Phi \to b\bar b+\tau\tau$ decays are so strongly
enhanced that they leave little room for other possibilities. Only in a few 
special situations that these SUSY decays can be significant.  For  the decays
$\Phi \to \chi \chi$, it is the case when  $i)$  all $\chi = \chi^\pm_{1,2}$ and
$\chi^0_{1\!-\!4}$ channels  are kinematically open or $ii)$ if only a subset of
$\chi$ particles is light, they should be mixtures of gauginos and higgsinos to
maximize the $\Phi \chi \chi$ couplings. Both  scenarios should be challenged by
the present LHC constraints\footnote{The  searches of charginos and neutralinos
in the same-sign lepton and  tri-lepton topologies at the LHC are now probing
significant portions of the gaugino--higgsino parameter space and they exclude
more and more the possibility of light $\chi$ states \cite{LHC-SUSY}. This is
particularly true   for mixed  gaugino--higgsino states in which  the $\Phi \chi
\chi$ couplings are maximised:  the lead to  a large gap between the lightest
and the next-to-lightest $\chi$ masses and hence a large  amount of missing
energy that make the searches more effective.}. In the case of sleptons, only
the decays into light $\tilde \tau$ states could be important; while the  decay
$A \to \tilde \tau_1 \tilde \tau_1$ is forbidden by CP--invariance, the decays
$H \to \tilde \tau_1 \tilde \tau_1$ and $H/A \to \tilde \tau_1 \tilde \tau_2$
can have substantial rates at high $\tb$  when the $\Phi \tilde \tau \tilde
\tau$ coupling is enhanced. However, again, at these large $\tb$ values,  the
$\Phi \to b\bar b$  and $\Phi \to \tau \tau$ decays are extremely enhanced and
leave little room for competition. 

Thus, only in the unlikely cases where the decay $H\! \to\! \tilde \tau_1 \tilde
\tau_1$ has a branching  rate of the order of 50\%, the squark loop contribution
to the $gg\to H$ process is of the order 50\%, or the $\Delta_b$ SUSY correction
is larger than 100\%, that one can change the   $pp \to \Phi \to \tau \tau$ rate
by $\approx 25\%$, which is the level  of the QCD uncertainty. One thus
expects $\sigma(pp \to \Phi)\times {\rm BR}(\Phi \to \tau  \tau)$ to be
extremely robust and to depend almost exclusively on $M_A$ and $\tb$.

Two more processes are considered for the heavier MSSM neutral Higgs bosons at
high $\tb$. The first one is  $pp \to \Phi \to \mu^+\mu^-$ for which the rate is
simply   $\sigma(pp\!\to\! \Phi \to \tau\tau )$ rescaled by BR$(\Phi \to \mu
\mu)/$BR$(\Phi \to \tau \tau)= m_\mu^2/m_\tau^2 \approx 4\times 10^{-3}$. The
rate is much smaller than in the  $\tau\tau$ case  and is not compensated by the
much cleaner $\mu \mu$ final state and the better resolution on the  invariant
mass. Searches in this channel have been performed in the SM Higgs case
\cite{LHC-mu} and the sensitivity is very low.   In addition, there is  the
process in which the $H/A$ bosons are produced in association with  two
$b$--quark jets and  decay into $b \bar b$ final states and searches in this
channel have been performed by the CMS collaboration  with the 7 TeV data
\cite{CMS-bbbb}. However, the  sensitivity is far lower than in the 
$\tau^+\tau^-$ channel.

Thus, the $pp \to \Phi \to \tau^+ \tau^-$ search for the neutral Higgs bosons 
provides the most stringent limits on the MSSM parameter space at large $\tb$ 
and all other channels are weaker in comparison and provide only cross checks.
We will thus concentrate on this process in the rest of our discussion of the
high  $\tb$ regime.

A final remark needs to be made on the charged Higgs boson. The dominant $H^\pm$
search channel at present energies is in  $H^\pm \to \tau \nu$  final states
with the $H^\pm$ bosons produced in top quark decays for masses not too close to
$M_{H^\pm}=m_t\!-\! m_b \approx 170$ GeV
\beq
pp \to t\bar t ~~{\rm with}~~ t \to H^+ b \to \tau\nu \; b
\eeq
This is particularly true at high $\tb$ values when the $t\to H^+b$ branching
ratio which  grows with $\bar m_b^2 \tan^2\beta$, is significant.  For higher
$H^\pm$ masses, one should rely on the three--body production  process $pp \to
tbH^\pm  \to tb \tau \nu$ which leads to a cross section that is also
proportional to $\tan^2\beta$, but the rates are presently too small. Hence,
processes beyond  $t\to bH^+$  can be considered only at the upgraded LHC.

\subsection{The low $\tb$ regime}

\begin{figure}[!h]
\begin{center}
\mbox{
\epsfig{file=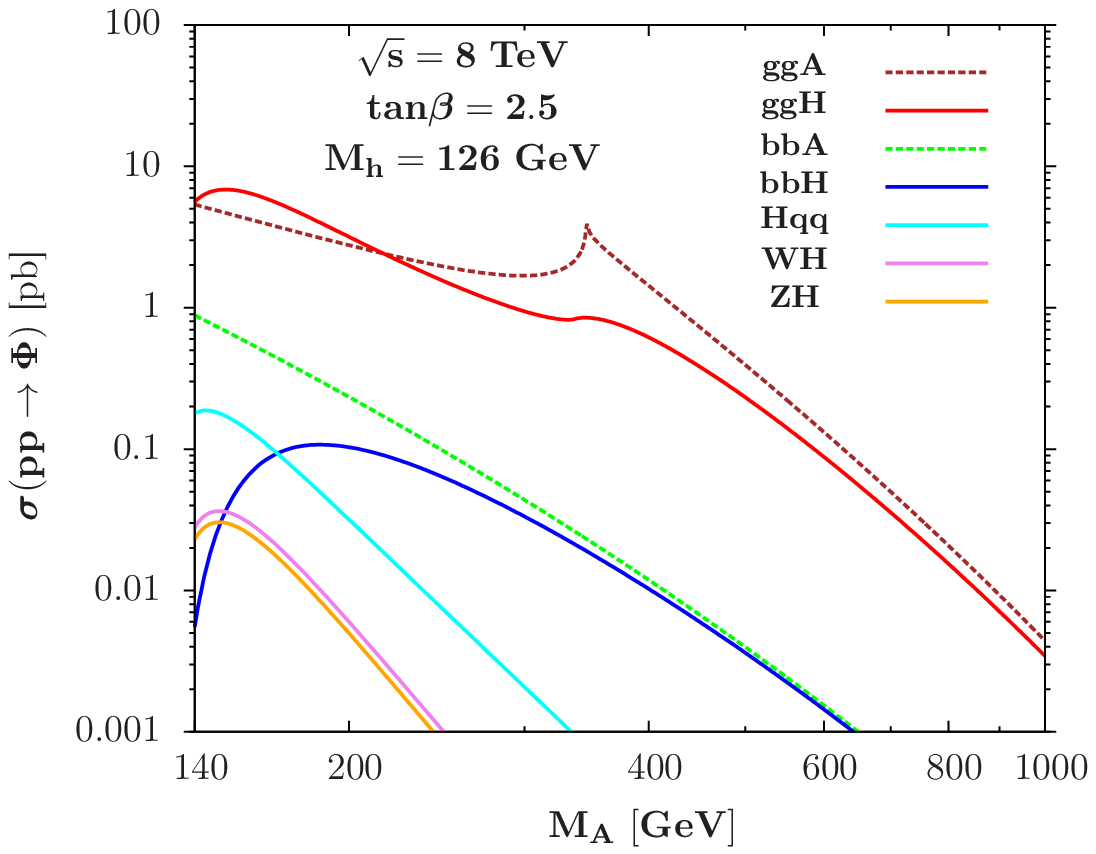,height=7cm,width=7cm} 
\epsfig{file=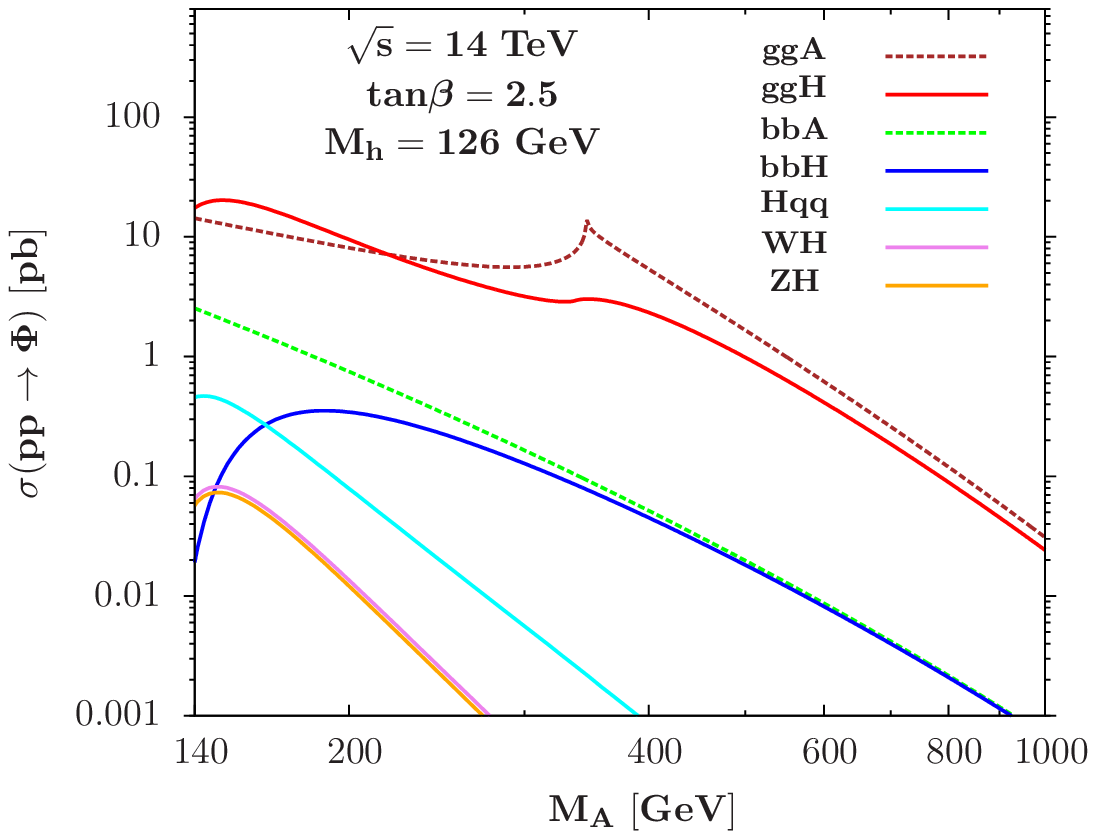,height=7cm,width=7cm} 
}
\end{center}
\vspace*{-5mm}
\caption[]{The production cross sections of the MSSM heavier Higgs bosons at
the LHC with $\sqrt s=8$ TeV (left) and $\sqrt s=14$ TeV (right) for $\tb=2.5$.
Only the main channels are presented. The higher order corrections are
included (see text) and the MSTW PDFs have been adopted.}
\vspace*{-1mm}
\label{Fig:xs}
\end{figure}

The phenomenology of the heavy MSSM $A,H,H^\pm$ bosons  is richer at low $\tb$
and  leads to a production and decay pattern that is slightly more involved than
in the high $\tb$ regime. Starting with the production  cross sections, we
display in Fig.~\ref{Fig:xs} the rates for the relevant $H/A/H^\pm$ production
processes at the LHC with center of mass energies of $\sqrt s=8$ TeV and  $\sqrt
s=14$ GeV assuming $\tb=2.5$. The programs {\tt HIGLU} \cite{HIGLU}, {\tt SUSHI}
\cite{SUSHI} and those of Ref.~\cite{Michael-web} have been modified in such a way
 that the radiative corrections in the Higgs sector are calculated 
according to section 2.1 and lead to a fixed $M_h=126$ GeV value. The  MSTW set
of parton distribution functions (PDFs) \cite{MSTW} has been adopted. For
smaller $\tb$ values,  the cross sections for the various processes, except for
$pp \to H/A+b\bar b$,  are even larger as the $H/A$ couplings to top quarks and
the $HVV$ coupling outside the decoupling limit are less suppressed. 

Because of CP invariance which forbids $AVV$ couplings, the pseudoscalar state
$A$  cannot be produced in the Higgs-strahlung and vector boson fusion
processes. For $M_A \gsim 300$ GeV, the rate  for the associated $pp \to t\bar t
A$ process is rather  small, as is also the case of the $pp \to b\bar b A$ 
cross section which  is not  sufficiently  enhanced by the $Abb \propto \tb$ 
coupling. Hence, only the $gg\to A$ fusion process with the dominant $t$--quark
and sub-dominant $b$--quark loop contributions included  provides large rates
at low $\tb$.

The situation is approximately the same for the CP--even $H$ boson: only the
$gg\to H$ process provides  significant production rates at relatively high
values of $M_H$, $M_H \gsim 300$ GeV,  and low $\tb$, $\tb \lsim 5$. As in the
case of $A$, the cross section for $pp \to t\bar t H$ is suppressed compared to
the SM case while the rate for   $pp \to b\bar b H$ is not enough enhanced.
However, in this case, the vector boson fusion $pp \to Hqq$ and Higgs-strahlung
processes $q\bar q \to HW/HZ$ are also at work  and have production rates that
are not too suppressed compared to the SM at sufficiently low $M_H$ values, 
$M_H \lsim 200$--300 GeV and $\tb \approx 1$.

Hence, for $M_A \gsim 300$ GeV, the only relevant production process is  $gg\to
\Phi$ with the dominant contribution provided by the heavy top quark loop. In
this case, one can include not only the large NLO QCD corrections
\cite{ggH-NLO-approx}, which are  known in the exact case \cite{ggH-NLO}, but 
also the NNLO QCD corrections \cite{ggH-NNLO} calculated in a effective
approach  with $m_t\! \gg\! M_\Phi$ which should work in practice  for
$M_\Phi \lsim 300$ GeV but can be extended to higher masses. 

For the charged Higgs boson, the dominant production channel in the low $\tb$
regime is again  top quark decays, $t \to H^+ b$, for $M_{H^\pm} \lsim 170$ GeV.
Indeed, for $\tb \lsim 5$, the  $m_t/\tb$ component of the $H^\pm tb$ coupling
becomes rather large, leading to a significant $t\to H^+b$ branching ratio. For
higher $H^\pm$ masses,  the main process to be considered is $gg/q\bar q  \to
H^\pm tb$ \cite{LO-H+}. As in the case of  $pp \to b\bar b\Phi$, one can take
the $b$--quark as a parton and consider the equivalent but simpler $2 \to 2$
channel $gb \to H^\pm t$. One obtains an accurate description of the cross
section if the renormalisation and factorisation scales are chosen  to be low,
$\mu_R=\mu_F \approx \frac16 (M_{H^\pm}+m_t)$ in order to account for the large
NLO QCD corrections \cite{NLO-H+}; the scales uncertainties are large though,
being of order 20\% \cite{LHCWG}. Additional sources of $H^\pm$ states for
$M_{H^\pm} \lsim 250$ GeV are provided by pair and associated production with
neutral Higgs bosons in $q\bar q$ annihilation as well as $H^+H^-$ pair and
associated $H^\pm W^\mp$ production in $gg$ and/or $b\bar b$ fusion  but the
rates are very small \cite{Stefano}.

\begin{figure}[t]
\begin{center}
\mbox{
\hspace*{-8mm}
\epsfig{file=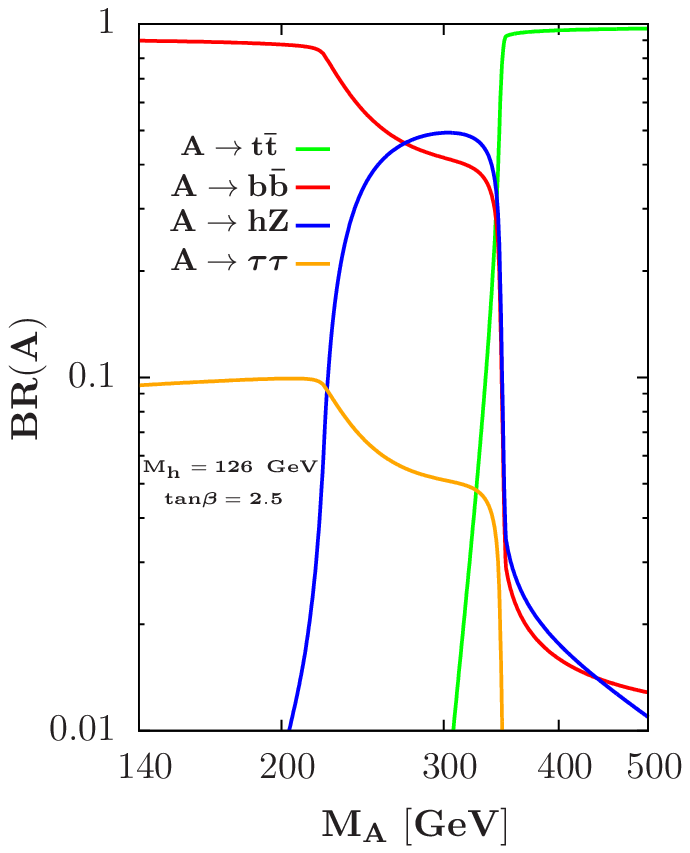,height=7cm,width=4.5cm}\hspace*{0mm}
\epsfig{file=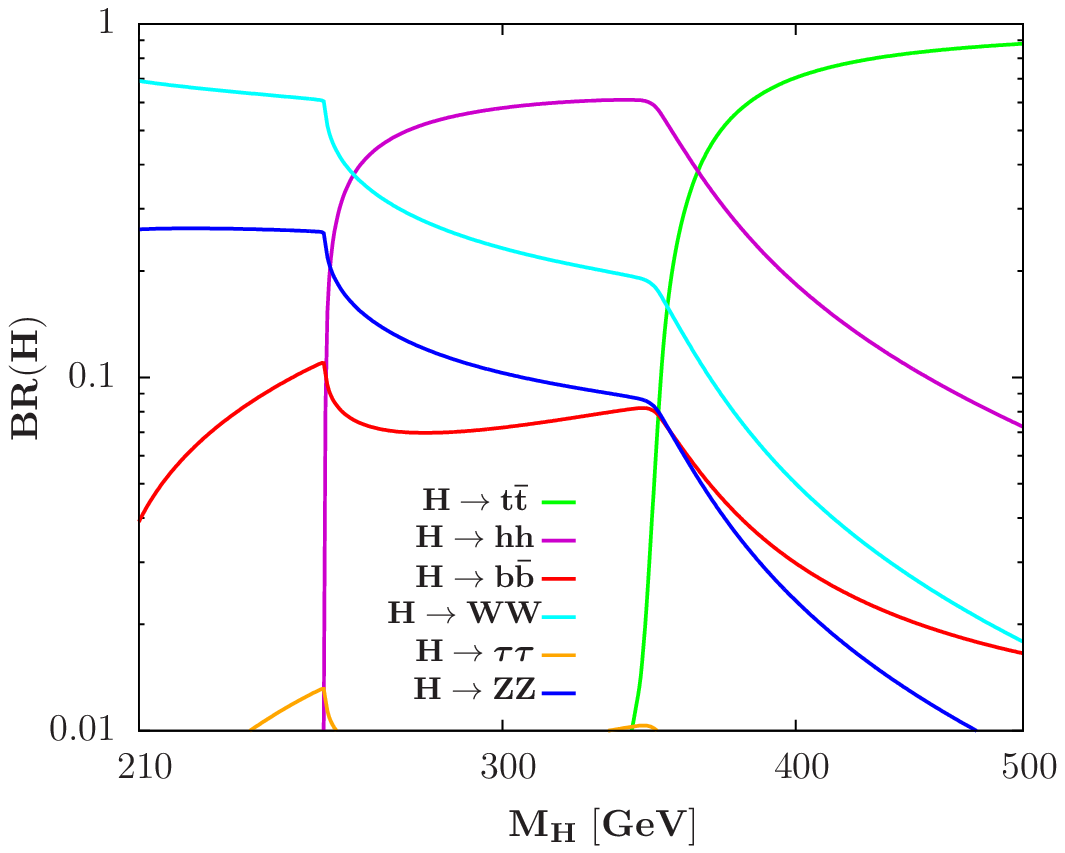,height=7cm,width=6.5cm}\hspace*{0mm} 
\epsfig{file=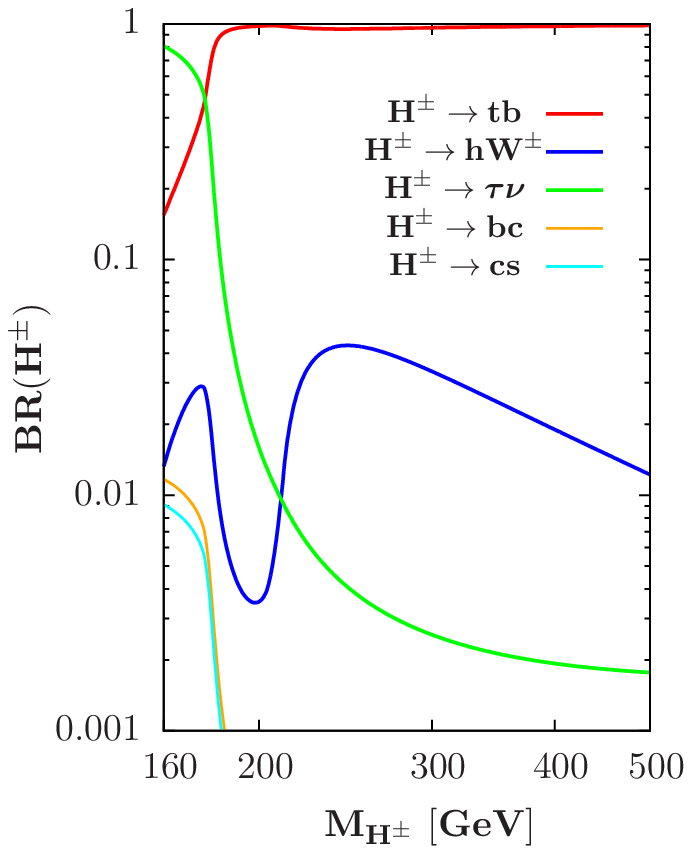,height=7cm,width=4.5cm} 
}
\end{center}
\vspace*{-5mm}
\caption[]{The decay branching ratios of the heavier MSSM Higgs bosons
$A$ (left), $H$ (center) and $H^\pm$ (right) as a function of their 
masses for  $\tb=2.5$. The program {\tt HDECAY} \cite{HDECAY} has been used
with modifications so that the radiative corrections lead to  
$M_h=126$ GeV.   }
\label{Fig:br}
\end{figure}

Let us turn to the decay pattern of the heavier MSSM Higgs particles which can
be rather involved  in the low $\tb$ regime. In this case, as the couplings of
the $H/A$ bosons to $b$--quarks are not very strongly enhanced and the couplings
to top quarks (and gauge bosons in the  case of the $H$ state)  not too
suppressed, many interesting channels appear. The branching fractions for the 
$H/A/H^\pm$ decays are shown in Fig.~\ref{Fig:br} as functions of their masses
at $\tb=2.5$. They have been obtained using the program {\tt HDECAY}
\cite{HDECAY} assuming large $M_S$ values that
lead to a  fixed $M_h=126$ GeV value. The pattern does not significantly depend
on other SUSY parameters, provided that Higgs decays into supersymmetric
particles  are kinematically closed as it will be implicitly assumed in the
following\footnote{In fact, even in this low $\tb$  case, the $t \bar t$ decays
for sufficiently large masses are so dominant that they do not lead to any
significant quantitative change if SUSY particles are light. In addition, being
not enhanced by $\tb$, the $\Delta_b$ correction has no impact in this low
$\tb$  regime.}, where the main features of the decays are summarised in a few
points.

-- Sufficiently above the $t\bar t$ threshold for the neutral and the $tb$
threshold for the charged Higgs bosons,  the decay channels $H/ A \rightarrow
t\bar{t}$  and $H^+ \to t \bar b$ become by far dominant for $\tb \lsim 3$ and
do not leave space for any other decay mode. Note that these decays have also
significant branching fractions below the respective kinematical thresholds
\cite{H3body}.  It is especially true for the charged Higgs state for which
BR($H^+\! \to\! t \bar b)\! \gsim \!1\%$ for $M_{H^\pm}\!\approx\! 130\;$GeV.  

-- Below the $t\bar t$ threshold, the $H$ boson can still decay into gauge
bosons $H \to WW$ and $ZZ$ with rather substantial rates as the $HVV$ couplings
are not completely suppressed.

-- In  the window $2M_h \lsim M_H \lsim 2m_t$, the dominant decay mode for $\tb
\lsim 3$ turns out to be the very interesting  channel  $H\to hh$ channel. As
discussed earlier, the $Hhh$ self--couplings given in  eq.~(\ref{Hhhcp}) is
significant at low $\tb$ values.

-- If allowed kinematically, i.e. for $M_A\! \gsim M_h+M_Z$ GeV, the CP--odd
Higgs boson can also decay into $hZ$ final states with a significant  rate 
below the $t\bar t$ threshold as the  $AZh$ coupling (that is the same as the
$HVV$ coupling) is substantial. Nevertheless, the $A \to \tau\tau$ channel   is
still important as it has a branching fraction above $\approx 5\%$ up to $M_A
\approx 2m_t$. 

-- In the case of the charged Higgs state, there is also the channel  $H^+ \to
Wh$  which is important similarly to the $A \to hZ$ case. Note that for $
M_{H^\pm} \lsim 170$ GeV, the decay $H^+ \to c\bar s$ that is usually considered
only in two--Higgs doublet models and the very interesting flavor  changing mode
$H^+ \to c\bar b$ have rates that are at the  percent level. All these exotic channels
have larger branching ratios, above $\approx 10\%$, for $\tb$ values close to
unity.

\subsection{The case of the h boson}

A final word is due to the production and decay rates of the lighter $h$ boson
that we will assume to be the state with a 126 GeV mass observed at the LHC.

\begin{figure}[!h]
\begin{center}
\mbox{
\epsfig{file=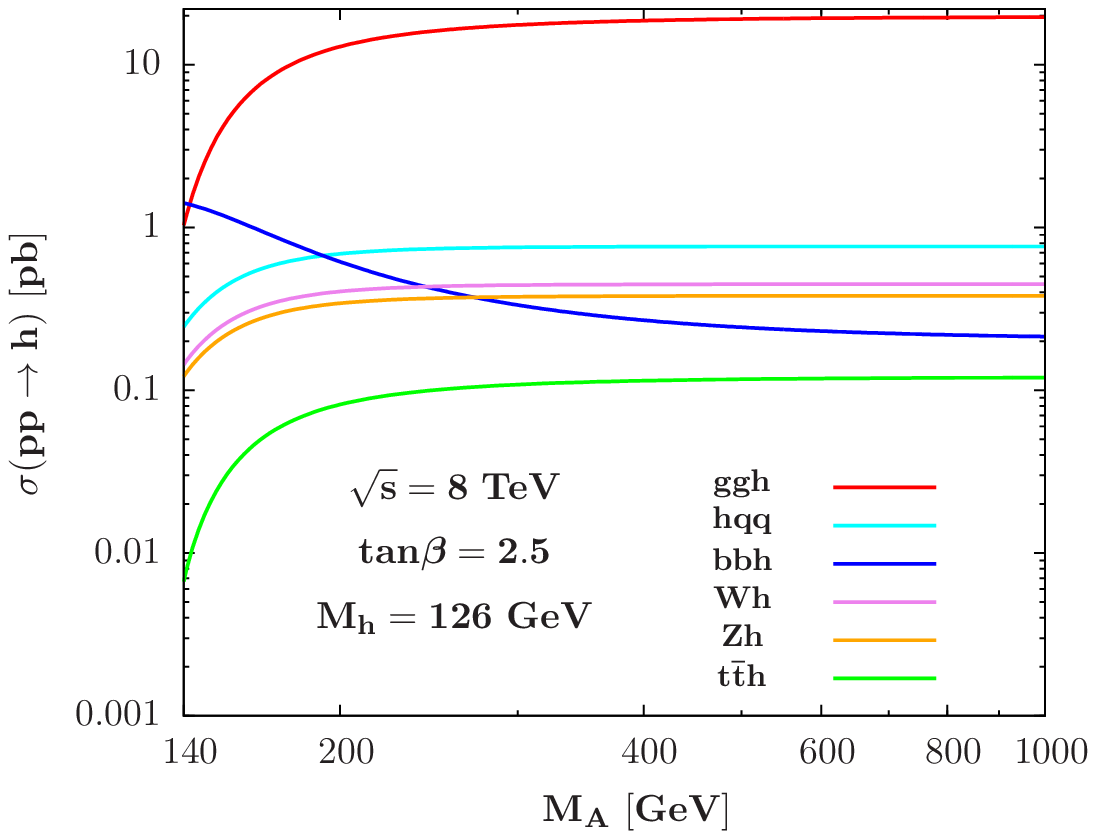,height=6.5cm,width=7cm}
\epsfig{file=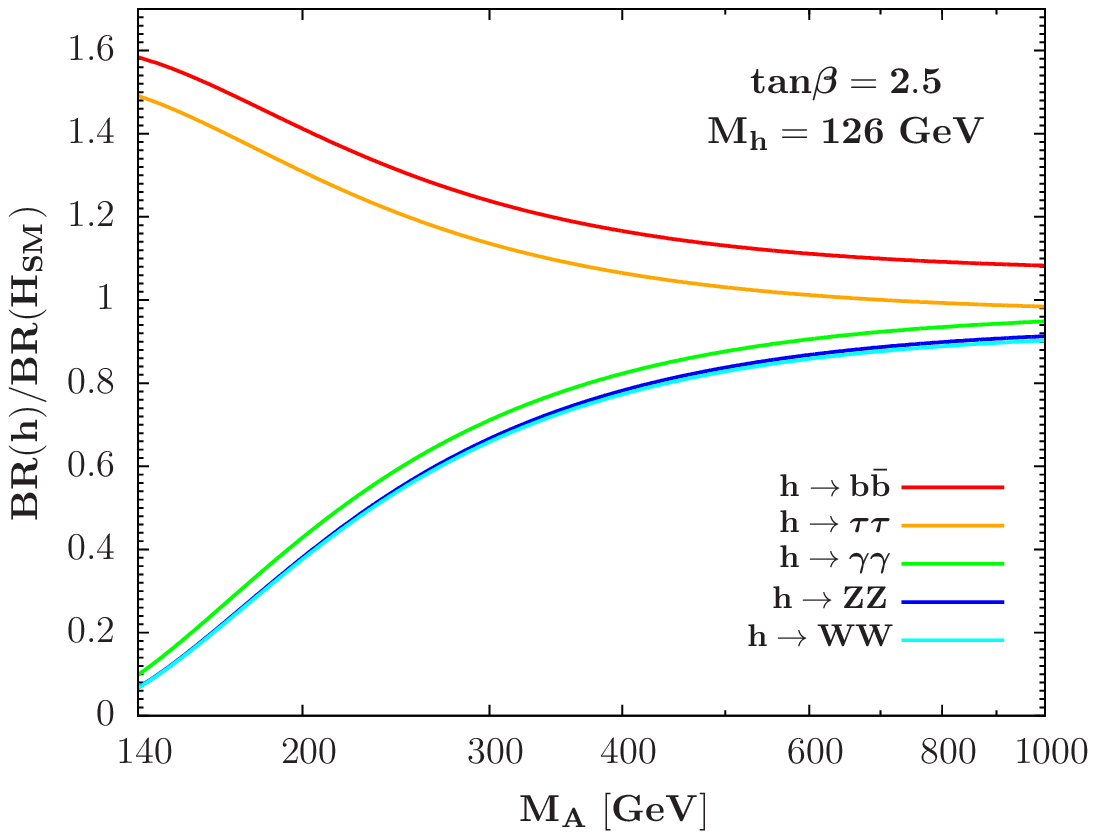,height=6.5cm,width=7cm} 
} 
\end{center}
\vspace*{-5mm}
\caption[]{The production cross sections of the lighter $h$ boson at
the LHC with $\sqrt s=8$ TeV (left) and the variation of its decay
branching fractions compared to the SM values (right) for $\tb=2.5$.
Again, the radiative corrections in the Higgs sector are such that 
$M_h=126$ GeV.}
\label{Fig:hboson}
\end{figure}

In the left--hand side of Fig.~\ref{Fig:hboson}, we display the cross sections
for the relevant Higgs production channels at the LHC with $\sqrt s=8$ TeV as a
function of $M_A$ at $\tb=2.5$.  Again, the radiative corrections in the $\epsilon$ 
approach are such that $M_h$ is fixed to 126 GeV.  Shown are the rates for the 
gluon fusion $gg \to h$, vector boson fusion $qq \to hqq$, Higgs--strahlung
$q\bar q \to hW, hZ$ as  well as associated $pp \to t \bar t h$ processes.  The
relevant higher order QCD corrections are implemented and  the MSTW set of PDFs
has been adopted. The rates can be very different whether one
is in the decoupling limit  $M_A \approx 1$ TeV where the $h$ couplings are
SM--like or at low $M_A$ values when the $h$ couplings are modified.

The variation of the branching ratios compared to their SM values, which
correspond to their MSSM values in the decoupling limit, are displayed as a
function of $M_A$ for $\tb=2.5$ in the right-hand side of the figure. Sown are
the branching fractions for the decays that are currently used to search for the
SM Higgs boson, i.e. the channels $h \to bb, \tau \tau, ZZ, WW, \gamma\gamma$. 
Again, large differences compared to the SM can occur at low to moderate $M_A$
values. 

The data collected so far by the ATLAS and CMS collaborations  on the observed
126 GeV Higgs particle should thus put strong constraints on the 
parameters $\tb$ and $M_A$.

\newpage

\section{Present constraints on the MSSM parameter space}

\subsection{Constraints from the $h$ boson mass and rates}

We start this section  by discussing the impact of the large amount of ATLAS and
CMS data  for the observed Higgs state at the LHC on the MSSM parameter space.
We will assume for definiteness that the $h$ boson is indeed the observed
particle as  the possibility that it is the $H$ state  instead is ruled out by
the LHC data \cite{paper3}.  

A first constraint comes from the measured mass of the observed state, $M_h
\approx 126$ GeV. As discussed previously and in several other instances such
as  Ref.~\cite{paper1}, in the phenomenological MSSM, this large $M_h$ value
indicates that the radiative corrections in the Higgs sector are maximised.  If
the
scale $M_S$ is close to 1 TeV as dictated by naturalness arguments, this implies
that one is in the decoupling regime (and hence, dealing with a SM--like Higgs
particle) with intermediate to high--$\tb$ values and maximal stop mixing. If
the  SUSY scale is pushed to $M_S \approx 3$ TeV, the highest acceptable value 
from fine-tuning adopted in  many analyses such that of
Refs.~\cite{paper2,paper3},  a smaller mixing in the Higgs sector and values of
$M_A$  of order of a few hundred GeV can be made possible. However, $\tan\beta$
values in the low regime, $\tb \lsim 3$--5 cannot be accommodated as  they lead to
$M_h \lsim 123$ GeV and even to $M_h\lsim 120$ GeV, which is the lowest value
that can be reached  when including the theoretical and the top-quark mass
uncertainties in the calculation of $M_h$.  

To obtain an acceptable value of $M_h$ in the low $\tb$ regime, one needs to
push $M_S$ to the 10 TeV domain or higher. In the approach that we are
advocating here, in which the radiative corrections in the MSSM Higgs sector are
implemented in the rather simple (but not completely inaccurate)  approximation
where only the leading RGE improved one--loop correction of eq.~(\ref{epsilon})
is taken into account, one can trade the (unknown) values of $M_S$ and the
mixing parameter $X_t$ with  the (known) value of the Higgs mass $M_h$. In other
words, for each set of $\tb$ and $M_A$ inputs, one selects the $\epsilon$
radiative correction that leads to the correct mass $M_h=126$ GeV. The LHC 
constraint on the mass of the observed Higgs state is then automatically
satisfied.  We emphasize again that for the large SUSY scales that are needed
for the low $\tb$ regime, $\tb \lsim 3$,  the MSSM spectrum cannot be calculated
in a reliable way using the usual versions of  the RGE programs  such as {\tt
Suspect}. 

A second constraint comes from the measurement of the production and  decay
rates of the observed Higgs particle. The ATLAS and CMS collaborations have 
provided the signal strength modifiers $\mu_{XX}$, that are  identified with
the  Higgs cross section times decay branching ratio normalized to the  SM
expectation  in a given $H\!\to\! XX$ search channel. For the various search
channels that have been conducted, $h \to ZZ, WW, \gamma \gamma, \tau\tau$ and
$b\bar b$ with the entire set of data collected in the runs at $\sqrt s=7$ TeV
and $8$ TeV, i.e. $\approx 5$ fb$^{-1}$ and $\approx 20$ fb$^{-1}$ (with the
exception of $h\to b\bar b$ which has been analyzed only with 17
fb$^{-1}$ of the 7+8 TeV data) numbers can be found in 
Refs.~\cite{ATLAS-gg,CMS-gg,ATLAS-ZZ,CMS-ZZ,ATLAS-WW,CMS-WW,ATLAS-bb,CMS-bb,ATLAS-tau,CMS-tau}.
These should be used to constrain the couplings of the $h$ state and, hence,
the  $[\tb, M_A]$ parameter space.  

Rather than performing a complete fit of the ATLAS and CMS light Higgs data
including all the signal strengths, we will simply use the most precise and
cleanest observable in this context: the signal strength $\mu_{ZZ}$ in the 
search channel $h\to ZZ$. As recently discussed in Ref.~\cite{ratios} (to which
we refer for the details), this channel  is  fully inclusive and  does not
involve the additional large theoretical uncertainties that occur when breaking
the cross section of the dominant production process  $gg\! \to\! h$  into jet
categories\footnote{For instance, the signal strengths in the  $ \tau\tau$ and
$WW$ channels are obtained by considering the $gg\to H+0j,1j$ and/or the vector
boson fusion  categories.  The signal strength $\mu_{WW}$ provides the same
information as $\mu_{ZZ}$, while the  measurement of the signal strengths in
the  $h\to b\bar b$ and $h\to \tau^+ \tau^-$ channels are not yet very accurate.
Hence, using only the $h\to ZZ$ channel should be a good approximation. }. In
addition, contrary to the global signal strength $\mu_{\rm tot}$, it does not
involve the channel $h\to \gamma\gamma$ which, at least in the ATLAS case,
deviates from the SM prediction and  might indicate the presence of new
contributions  (such as those of light charginos?)  in the  $h\gamma\gamma$
loop. The combination of the ATLAS and CMS data in the $ZZ$ channel gives, 
$\mu_{\rm ZZ} =1.10 \pm 0.22 \pm 0.20$   where the first uncertainty is
experimental and the second one theoretical.    Following Ref.~\cite{Baglio2},
we assume a total theoretical uncertainty of $\Delta^{\rm th}\!=\!\pm 20\%$ and,
since it should be considered  as a bias,   we  add it linearly to the
experimental error. This gives a lower limit on the $h\to ZZ$ signal strength of
$\mu_{ZZ}\gsim 0.62$ at  68\%CL and  $\mu_{ZZ}\gsim 0.4$ at  95\%CL. 

In the MSSM case,  the signal strength will be given by $\mu_{ZZ}\!=\! \sigma
(h)\! \times \!{\rm BR}(h \!\to \!ZZ)/$ $\sigma (H_{\rm SM})  \times {\rm BR}(H_{\rm SM}
\to ZZ)$ and  will be thus proportional  to combinations of reduced $h$ coupling
squared to fermions and gauge bosons, $g_{htt}^2 \times g_{hVV}^2 /
g_{hbb}^2$... The fact that $\mu_{ZZ}$ can be as low at 0.4 at 95\%CL means that
we can  be substantially far from the decoupling limit, $g_{HVV}^2 \approx 0.1$,
with not too heavy $H/A/H^\pm$  states even at low $\tb$.

\begin{figure}[!h]
\begin{center}
\vspace*{-3mm}
\epsfig{file=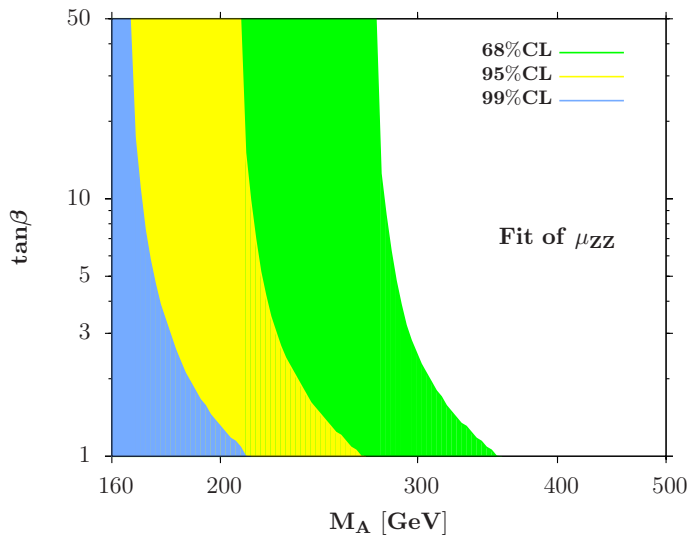,height=7cm} 
\vspace*{-5mm}
\end{center}
\caption[]{The $[\tb, M_A]$ parameter space of the MSSM in which the signal 
strength in the $h\to ZZ$ search channel is not compatible with the LHC data on
the rates of the observed $h$ boson at the 68\%CL (green), 95\%CL (yellow)
and 99\%CL (blue).}
\label{Fig:scan-Mh}
\end{figure}

In Fig.~\ref{Fig:scan-Mh}, we have scanned the $[\tb, M_A]$ parameter space and
delineated the areas in which the 68\%CL and 95\%CL constraints on  $\mu_{ZZ}$
are fulfilled.  We observe that indeed, the entire range with $M_A \lsim 200$
GeV for most value of $\tb$ is excluded at the 95\%CL. With increasing $\tb$,
the excluded $M_A$ values are lower and one recovers the well known fact that 
the decoupling limit is reached more quickly  at higher $\tb$ values. In most
cases, we will use this indirect limit of $M_A \lsim 200$ GeV  prior to any
other constraint (except for illustrations in the $H^\pm$ case where the low
mass range will be kept). 

\vspace*{-10mm}
\subsection{Constraints from the heavier Higgs searches at high $\tb$}

As discussed in section 3.1,   the most efficient channel to search for the
heavier MSSM Higgs bosons is by far $H/A$ production in $gg$ and $b\bar b$
fusion with the Higgs bosons decaying  into $\tau$ lepton pairs, $pp \to \Phi
\to \tau^+ \tau^-$.  Searches for this process have been performed by the  ATLAS
collaboration with $\approx 5$ fb$^{-1}$ data at the 7 TeV run \cite{ATLAS-tau}
and by the CMS collaboration with $\approx 5+12$ fb$^{-1}$ data at the 7 TeV and
8 TeV runs \cite{CMS-tau}. Upper limits on the production times decay rates of
these processes (which, unfortunately,  have not given by the collaborations)
have been set and they can be turned into constraints on the MSSM parameter
space which, in the Higgs sector, corresponds to the $[\tb,M_A]$ plane. 

\begin{figure}[!h]
\begin{center}
\vspace*{-2mm}
\epsfig{file=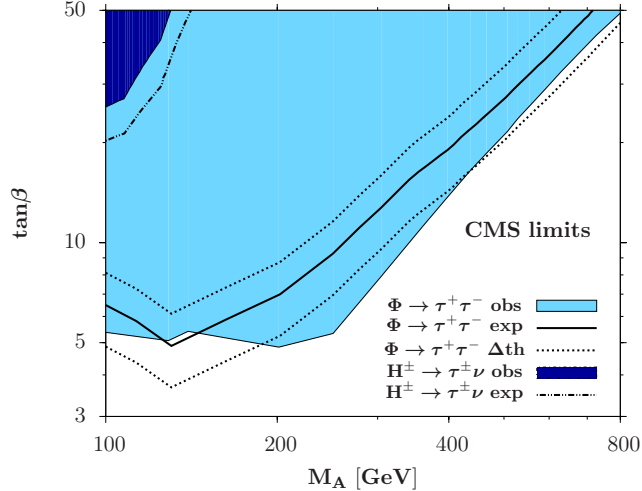,height=6.5cm} 
\vspace*{-2mm}
\caption{The $[\tb,M_{A}]$ plane in the MSSM  in which the $pp \to H/A
\to  \tau^+\tau^-$  (light blue) and $t \to bH^+ \to b \tau \nu$ (dark blue) 
search constraints using the CMS data are included (observed limits). The 
solid contour for the $pp\to \tau \tau$ mode is for the median expected limit 
and the two dashed ones are when the QCD uncertainties on the rates are 
included.}
\label{fig:MA-tb}
\vspace*{-5mm}
\end{center}
\end{figure}

In Fig.~\ref{fig:MA-tb}, we display the sensitivity of the CMS $\Phi \to \tau
\tau$ analysis   in the $[\tb,M_A]$ plane. The excluded  region,
obtained from the observed limit at the 95\%CL  is drawn in light blue. The
solid line  represents the median expected limit which turns out to be weaker
than the observed limit.  As can be seen,   this constraint is extremely
restrictive and for values $M_A \lsim 250$ GeV, it excludes  almost the entire
intermediate and high  $\tb$ regimes, $\tb \gsim 5$.   The constraint is of
course less effective for a heavier  pseudoscalar Higgs boson, but even for
$M_A=400$ GeV the high $\tb \gsim 10$ region is excluded and one is even
sensitive to large values  $M_A \approx 700$ GeV for  $\tb \gsim 50$. 

There are, however, some caveats to this exclusion limit as discussed in 
section 3.1. The first one is that  there is a theoretical uncertainty  that
affects the Higgs production cross section times decay branching ratios which is
of the order of $\pm 25\%$ when the $gg \to \Phi$ and $b\bar b \to \Phi$ cross
sections   are combined.  If this theoretical uncertainty is included when
setting the limit in the $[\tb,M_A]$ plane, as shown by the dashed contours 
around the expected limit in Fig.~\ref{fig:MA-tb}, the constraint will be
slightly weaker as one then needs to  consider the lower value of the  $\sigma(p
p \! \to \! \Phi) \! \times \! {\rm BR}(\Phi \to \tau^+ \tau^-)$ rate predicted
by theory. 

The second caveat is that the CMS (and ATLAS) constraint has been given in a
specific benchmark scenario, the maximal mixing scenario with the choice
$X_t/M_S= \sqrt 6$ and the value of the SUSY scale set to $M_S=1$ TeV; the other
parameters such as the gaugino and higgsino masses and the first/second
generation fermion parameters that have little impact  can be chosen as in
eq.~(\ref{pbenchmark}). However, as was previously argued,  the $pp \to \Phi \to
\tau \tau$ cross section times decay branching fraction is very robust and,
hence, the  exclusion limit  is almost model independent.  It is altered  only
very mildly by the radiative corrections in the  MSSM Higgs sector, in
particular by the choice of the parameters $M_S$ and $X_t$ (this is especially
true if these parameters are to be traded against the measured values of $M_h$).   

In fact, the exclusion limit in Fig.~\ref{fig:MA-tb} can be obtained in any MSSM
scenario with the only assumption being that  SUSY  particles are too heavy to
affect $\sigma(pp\to \Phi) \times {\rm BR} (\Phi \to \tau\tau)$ by more than 25\%,
which is the estimated theoretical uncertainty. Even in the case of light SUSY
particles, it is very hard to make that stop/sbottom squarks contribute
significantly to the  $gg \to H$ production processes, or to have a significant 
$\Delta_b$ correction to the $\Phi bb$ coupling which largely cancels out as
indicated by eq.~(\ref{deltab}), or to have a substantial  change of the $\Phi
\to \tau\tau$  fraction due to light SUSY particles  that appear in the
decays. 

Thus, the limit for the $pp \to \tau^+ \tau^-$ searches  is robust with respect
to the SUSY parameters and  is valid in far more situations and scenarios  than
the  ``MSSM $M_h^{\rm max}$ scenario" that is usually quoted by the experimental
collaborations. We thus suggest to remove this assumption on the benchmark
scenario (in particular it adopts the choice $M_S=1$ TeV which  does not allow 
low $\tb$ values and which starts to be challenged by direct SUSY searches), as
the only relevant assumption, if any, should be that we do not consider cases in
which the SUSY particles are too light to alter the Higgs production and decay
rates.  This is  a very reasonable attitude if we are interested mainly in the
Higgs  sector.

Another constraint on the MSSM Higgs sector\footnote{A search has also been
performed by  the CMS collaboration  based on the 7 TeV data in the channel  
$pp \to \Phi b\bar b \to  bbbb$ \cite{CMS-bbbb}. This search  is much less
sensitive than the $\tau\tau$ search even if one extrapolates the expected 
limits to the same amount of data. We will thus ignore it in our study.} is the
one from charged Higgs searches in the $H^- \to \tau \nu$  final states with the
$H^\pm$ bosons produced in top quark decays, $t \to H^+ b \to \tau\nu  b$.  Up
to now, the ATLAS and CMS collaborations have released results only with the
$\approx 5$ fb$^{-1}$ collected at $\sqrt s=7$ TeV \cite{ATLAS-H+,CMS-H+}.  We
have also delineated in Fig.~\ref{fig:MA-tb} the impact on the  $[\tb,MA]$
parameter space  of the CMS 95\%CL observed limits in this channel. 

As can be observed, the constraint is effective only for values $M_A \lsim 150$
GeV  which correspond to a light $H^+$ state that could be produced in top quark
decays.  The search is sensitive to the very  high $\tb$ region which is
completely excluded by the $\tau \tau$ search, that is performed with much more
data though. However, even if the comparison is made for the same amount of
data, he $pp \to \Phi \to \tau\tau$ search is by far more sensitive.    

Note that contrary to the $pp\to \tau^+ \tau^-$ case, the limits at high $\tb$
from the process $pp \to t\bar t$ with  $t \to bH^+ \to b  \tau \nu$  might be
more model dependent. Indeed, while SUSY decays  might not be important as the
small $M_{H^\pm}$ value leaves little room for light sparticles (and the high
$\tb$ values would suppress these decays anyway), the effect of the $\Delta_b$
corrections might be larger as there is no cancellation between production and
decay rates. Nevertheless, the $H^\pm$ limit is effective only for $M_A \lsim
150$ GeV values excluded by the  $h$ data. We  keep this $H^\pm$ constraint
though, as it is also valid in two-Higgs  doublet models.

\subsection{Extrapolation to the low $\tb$ region and the full 7+8 data}

A very important remark is that in our version of the constraints in the $[\tb,
M_A]$ plane of Fig.~\ref{fig:MA-tb}, we have removed the region excluded by the
bound on the $h$ mass, $M_h \gsim 114$ GeV from negative Higgs searches at LEP2,
which is also usually displayed by the experimental collaborations. In the usual
benchmark scenario, this constraint excludes the entire low $\tb$ regime,  $\tb
\lsim 3$, and at low $M_A\approx 100$ GeV, $\tb$ values up to $\tb \approx
10$.  

 A first reason for removing the ``LEP exclusion" region is that it is now
superseded by the ``observation" constraint $123~{\rm GeV} \lsim  M_h \lsim
129~{\rm GeV}$ (once the theoretical and experimental uncertainties are
included) which is by far stronger. In fact, as was discussed in
Ref.~\cite{paper2}, if the benchmark scenario with  $M_S=1$ TeV and maximal stop
mixing is to be adopted, the entire range  $\tb \lsim 5$ and $\tb \gsim 20$ for
any $M_A$ value would be excluded simply by requiring that  $123~{\rm GeV}
\lsim  M_h \lsim 129~{\rm GeV}$ (and the excluded regions would be completely
different for other $M_S$ and $X_t$ values as also shown in
Ref.~\cite{paper2}).  

A second reason is that the LEP2 $M_h$ constraint and even the constraint $M_h
\gsim123$ GeV  can be simply evaded for any value of $\tb$ or $M_A$ by
assuming  large enough $M_S$ values as discussed in section 2.1.  This will then
open the very interesting low $\tb$ region which can be probed in a model
independent way by Higgs search channels involving the $H,A,H^\pm$ bosons,
including the $t \to bH^+ \to b \tau \nu$ channel discussed previously.  

\begin{figure}[!h]
\begin{center}
\vspace*{-3mm}
\epsfig{file=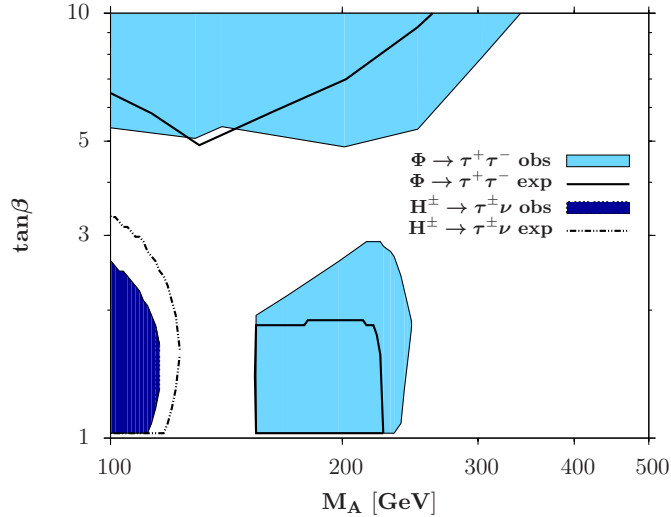,height=6.8cm} 
\vspace*{-2mm}
\caption{The $[\tb,M_{A}]$ plane in the MSSM  in which the $pp \to H/A
\to  \tau^+\tau^-$  (light blue) and $t \to bH^+ \to b \tau \nu$ (dark blue) 
observed limits using the CMS data are extrapolated to low $\tb$. The 
solid contours in the $\tau\tau$ and $\tau\nu$ cases are for the expected limits.}
\label{fig:MA-tb-low}
\vspace*{-7mm}
\end{center}
\end{figure}

Indeed, the branching fraction for the decay $t \to bH^+$ is also significant at
low $\tb$ values, when the component of the  coupling $g_{tbH^+}$ that is
proportional to $\bar m_t /\tb$ becomes dominant. On the other hand, the
branching fraction for the decay $H^\pm \to \tau \nu$ stays close to 100\%. 
Hence, the rates for $pp \to t\bar t$ with $t \to bH^+ \to b\tau \nu$ are 
comparable for $\tb \approx 3$ and $\tb \approx 30$ and the processes can also 
probe the low $\tb$ region. This is exemplified in Fig.~\ref{fig:MA-tb-low}
where the $t \to bH^+$ CMS median expected and observed  limits obtained with
the 7 TeV data are extrapolated to the low $\tb$ region. As can be seen, the
region $\tb \lsim 2$ is excluded for $M_A\! \lsim\! 140$ GeV (this region
can also be probed in the $H^+\! \to\!  c\bar s$ mode). 

In fact, as is shown in the lower part of Fig.~\ref{fig:MA-tb-low}, even the
channel $pp \to \Phi \to \tau \tau$ is useful at low $\tb$. Indeed, for $\tb$
values close to unity, while the $b \bar b \to \Phi$ process becomes irrelevant,
the cross sections for the $gg \to \Phi$ process becomes very large, the reason
being that for $\tb \approx 1$ the couplings $g_{\Phi tt} \propto \bar m_t/\tb$
are significant and the  dominant top quark loop contribution becomes less
suppressed compared to the SM.  On the other hand, at least in the case of the
pseudoscalar $A$, the branching ratio for the $\tau^+\tau^-$ decay stays
significant for $M_A$ values up to the $t\bar t$ threshold as shown in 
Fig.~\ref{Fig:br}. Hence, the production times decay rate for $gg\to A \to
\tau\tau$ stays large and the CMS search limit is effective and excludes $\tb$
values close to 1, for pseudoscalar masses up to $M_A \approx 350$ GeV. 

\begin{figure}[!h]
\begin{center}
\vspace*{-5mm}
\epsfig{file=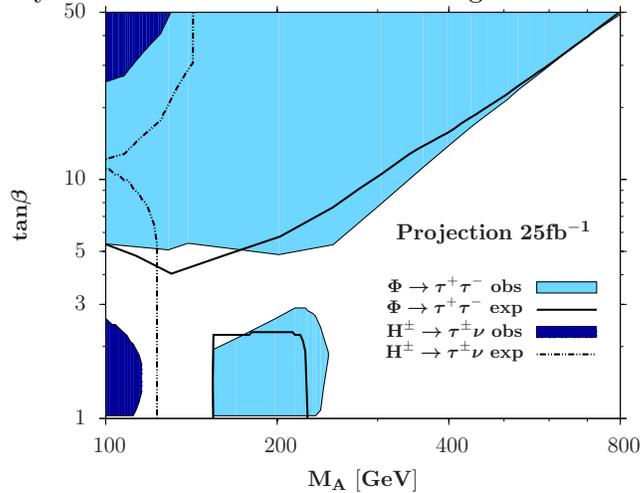,height=6.5cm} 
\vspace*{-2mm}
\caption{The $[\tb,M_{A}]$ plane in the MSSM  in which the $pp \to H/A
\to  \tau^+\tau^-$  and $t \to bH^+ \to b \tau \nu$ CMS expected limits 
are extrapolated to the full 7+8 TeV data with $\approx 25$ fb$^{-1}$.
The present observed limits are still shown in blue.}
\label{fig:MA-tb-comb}
\vspace*{-6mm}
\end{center}
\end{figure}

One would get a better feeling of the power of these constraints at low $\tb$
values (and in the charged Higgs  case also at high $\tb$), if the present
limits  in the $pp \to \tau \tau$ and $t \to bH^+ \to b \tau \nu$ channels are
extrapolated  to the full set of data collected in the 2011 and 2012 LHC runs.
This is shown in Fig.~\ref{fig:MA-tb-comb} where the median expected CMS limits
in the two search channels are extrapolated to an integrated luminosity of 25
fb$^{-1}$, assuming  that the limits simply scale like the square--root of the
number of events.  

The gain in sensitivity is very significant in the $H^\pm$ case  as the gap
between the present CMS limit with the $\approx 5$ fb$^{-1}$ of the 7 TeV data
and the expected limit with the additional $\approx 20$ fb$^{-1}$ data  at 8 TeV
is large (there is an additional increase of the  $pp \to t\bar t$ production
cross section from $\sqrt s=7$ TeV to 8 TeV). In the case of the $pp \to \tau
\tau$ channel, the increase of sensitivity is much more modest, not only because
the gap from the 17 fb$^{-1}$ data used in the latest CMS analysis and the full
25 fb$^{-1}$ data collected up to now is not large but, also, because  presently
the observed limit is much stronger than the expected limit.

Hence, these interesting low $\tb$ areas that were thought to be buried under
the LEP2 exclusion bound on $M_h$  are now open territory for heavy MSSM Higgs
hunting. This can be done not only in the two channels $pp \to \tau^+\tau^-$ and
$ t \to bH^+ \to b \tau \nu$ above (and which were anyway used at high $\tb$)
but also  in a plethora of channels that have not been discussed before (or at least 
abandoned after the LEP2 results)  and to which we turn now.  

\section{Heavy Higgs searches channels at low $\tb$}

We come now to the main phenomenological issue of this paper: the probe at the
LHC of the low $\tb$ region for a not too heavy pseudoscalar $A$ 
state\footnote{This issue has been discussed in the past and a summary can be
found in section 3.3.2 of Ref.~\cite{Review2}. It has been also addressed 
recently in Ref.~\cite{ABM} (where, in particular, a feasibility study of the
$H\to hh$ and $A\to hZ$ modes at $\sqrt s\!=\!14$ TeV is made) and in talks
given the last months by one of the authors \cite{talks}. Recents analyses of
heavier MSSM Higgsses  at intermediate and high $\tb$ can be found in
Refs.~\cite{Maiani:2012ij,others}.}.  We stress again that  this region can be resurrected 
simply  by allowing a large SUSY scale $M_S$ which removes the LEP2 $M_h \gsim
114$ GeV constraint (and now the LHC mass constraint $M_h \approx 126$ GeV).  We
show that several  channels discussed in the case of a high mass SM Higgs or in
scenarios beyond the SM can be used for the search of the MSSM $H,A$ and $H^\pm$
bosons.  

\subsection{The main search channels for the neutral H/A states}

\subsubsection{The $\mathbf{H \to WW,ZZ}$ channels}

These  are possible only for the heavier $H$ boson (because of CP invariance
there are no VV couplings for $A$) with masses below the $t \bar t$ threshold
where the branching  ratios for the decays $H\to WW$ and $H\to ZZ$ are
significant; see  Fig.~\ref{Fig:br}.   The $H\to WW$ process is particularly
useful in the region $160 \lsim  M_H \lsim 180~{\rm GeV}$ where the branching
ratio is close to 100\%. In both cases, the $gg \to H$ production process can
be used but, eventually,   vector boson fusion  can also be relevant
at the lowest $\tb$ and $M_H$ possible values. 

The search modes that are most useful at relatively low $M_H$ values,  should
be  the $pp \to H\to ZZ \to 4\ell^\pm$ and $pp \to H \to WW \to 2\ell 2\nu$
channels that have been used to observe the SM--like light $h$ boson (as the
mass resolution of the $H\to  WW$  channel is rather poor, one has to subtract
the observed signal events in  the low mass range, $M_H \lsim 160$ GeV) and to
exclude a SM--like Higgs particle with a mass up to 800 GeV
\cite{CMS-WW,CMS-ZZ}.  When the two processes are combined,  the  sensitivity is
an order of magnitude larger than for the SM Higgs for  masses below 400 GeV and
one can thus afford a substantial reduction of the couplings $g_{Htt}$ and
$g_{HVV}$ which should allow to probe  $\tb$ values significantly higher than  
unity\footnote{The ATLAS collaboration has recently analyzed  heavy $H$
production in a two--Higgs doublet model in the channel $H\to WW \to e\nu \mu
\nu$ with 13 fb$^{-1}$ data collected at $\sqrt s=8$ TeV \cite{ATLAS-2HDM}. 
Unfortunately, this analysis cannot readily be used as  the limit on the cross
section times branching fraction has not  been  explicitly given and the results
are displayed in terms of $\cos(\alpha)$ (and not  $\cos(\beta-\alpha)$ which
would have corresponded to the $HWW$ coupling) which does not allow an easy
interpretation in the MSSM.}. At high $H$ masses, $M_H \gsim 300$ GeV, one could
also add  the $pp \to H \to ZZ \to 2\ell 2q, 2\nu 2q, 2\ell 2 \nu$  and $pp \to
H \to WW \to \ell \nu 2q$  channels to increase the statistics, as done
in a recent study by the CMS collaboration  \cite{CMS-VV}. 

There is one difference with the SM Higgs case though. While in the SM,  the
Higgs particle has a large total width at high masses as a result of the decays
into longitudinal $W/Z$ bosons which make it grow as $M_{H_{\rm SM}}^3$, the
MSSM $H$ boson remain  narrow as the coupling $g_{HVV}$ is suppressed. In fact,
all MSSM Higgs particles will  have a total width that is smaller than $\approx
3$ GeV for $\tb  \approx 3$ and masses below 500 GeV. The smaller total width
in the MSSM can be rather helpful at relatively high $H$ masses as, for
instance, it allows to  suppress the continuum $ZZ$ background  by selecting
smaller bins for the invariant mass of the $ZZ$ system in the signal events. 
Issues like the interference of the signal  and the $gg \to VV$ backgrounds will
also  be less important than in the SM. 

\subsubsection{The $\mathbf{H/A \to t\bar t}$ channels}

This search channel has not been considered in the case of the SM Higgs boson
for two reasons~\cite{Review1}. The first one is that for $M_{H_{\rm SM}} \gsim
350$ GeV, the $H_{\rm SM} \to WW,ZZ$  channels are still relevant and largely
dominate over the $ H_{\rm SM}\to t\bar t$ decay channel which has a branching
fraction that is less than 20\% in the entire Higgs mass range (the reason being
again that the partial widths for $ H_{\rm SM} \to VV$ grow as $M_{H_{\rm
SM}}^3$ while for $H_{\rm SM} \to t\bar t$ it grows only like $M_{ H_{\rm
SM}}$). The other reason  is that the continuum $t\bar t$ background was
thought to be overwhelmingly large as it had to be evaluated in a large mass
window because of the large Higgs total width (in addition, the events from
$H_{\rm SM} \to t\bar t$ produce a dip--peak structure in the $gg\to t\bar t$
invariant mass spectrum that was unobservable for a large total width). 

The situation in the MSSM is very different. First, as mentioned previously, the
total width for heavy $H$ and $A$ states are much smaller, less than $\lsim  20$
GeV for any $\tb \gsim 1$ value for $M_{H,A} \lsim 500$ GeV and grow (almost)
linearly with the Higgs masses beyond this value. One can thus integrate the
$t\bar t$ continuum background in a smaller invariant mass bin and significantly
enhance the signal to background ratio. A second feature is that contrary to the
SM case,  the branching ratios for the  $H/A \to t\bar t$ decays  are almost
100\% for $\tb \lsim 3$ as soon as the channels are kinematically open (this is
particularly true for $A$ where even below the threshold, the
three--body decay  $A \to tt^* \to tbW$ is important). 

The only disadvantage compared to the SM is that the production cross section
could be smaller. In the MSSM, the only relevant process in the low $\tb$ regime
for $M_\Phi \gsim 350$ GeV is $gg \to \Phi$ with the dominant (almost only)
contribution being due to the top quark loop. The latter is suppressed by the
square of the coupling $g_{\Phi tt} \propto 1/\tb$ if $\tb$ is not close to
unity. However,  in the MSSM, one has to add the cross sections of both the $H$
and $A$ states. In addition,  the loop form factors  in the pseudoscalar $A$ and
scalar $H/H_{\rm SM}$ cases are different and,  as can been seen from
Fig.~\ref{Fig:xs},  the $gg \to \Phi$ cross section is larger in the
pseudoscalar Higgs case when  the same top Yukawa coupling is assumed.  

In toto, the situation for $H/A \to t\bar t$ will certainly be more favorable
for the MSSM at low $\tb$ than in the SM. While there was no search for the  SM
Higgs in this channel, the ATLAS \cite{ATLAS:1207.2409} and CMS
\cite{CMS:1211.3338} collaborations have looked  for heavy resonances (such as
new $Z'$ gauge bosons in extended gauge models or Kaluza--Klein excitations in
scenarios with  extra space--time dimensions) that decay into $t\bar t$ pairs
with the  data collected at the 7 TeV run.  The lepton+jets final state has been
studied in the topology where the top quarks are highly boosted which allows  a
good discrimination from the continuum $t\bar t$ background \cite{ttH-boost} 
(the ATLAS and CMS collaborations searches assume resonance masses $M_{tt} \gsim
700$ GeV to benefit from this topology).  Limits on the cross sections times
branching ratios have been set,  corresponding to roughly $\sigma_{tt} \approx
0.7$ pb for a resonance with a mass of $1$ TeV and  a narrow width, $\Gamma_{tt}
\approx 10^{-2} M_{tt}$ (which is more or less the case of the MSSM $H/A$ states
at $\tb \approx 3$). A lower (higher) cross  section is needed at larger
(smaller) resonance mass when the top quarks are (not) sufficiently boosted and,
at $M_{tt} \approx 500$ GeV, one needs  $\sigma_{tt} \approx 3$ pb which
approximately  corresponds to an increase with $1/M_{tt}^2$.

\subsubsection{The $\mathbf{A \to Zh}$ channel}

As discussed earlier, the $gg \to A$ production cross section is very large at
low $\tb$ values: it is higher than for the SM Higgs boson at $\tb =1$  (as the
form factor for the $ggA$ amplitude is larger than in the scalar Higgs case) and
is suppressed only by a factor $g_{Att}^2 \propto 1/{\rm tan}^2\beta$. On the
other  hand, in the range $M_h + M_Z \lsim M_A \lsim 2m_t$, the branching ratio
for the decay $A \to hZ$ is large for $\tb \approx 3$ and largely dominant  for
$\tb \approx 1$. In the mass window $M_A=210$--350 GeV, the production times
decay rate for the process $gg \to A \to hZ$ should be thus very high in the low
$\tb$ region. 

The $hZ$  final state has been searched for in the SM in the Higgs--strahlung 
process, $q \bar q \to Z^* \to Zh$ with the $Z$ boson decaying into leptons or
neutrinos, $Z\to \ell^+ \ell^-, \nu \bar \nu$ and the $h$ boson decaying into 
$b \bar b$ final states \cite{ATLAS-bb,CMS-bb}. The significance of the signal
is strongly increased by looking at boosted jets when the Higgs has a large
transverse momentum \cite{VH-boost}. In the CMS analysis with 17 fb$^{-1}$ of
the 2011 and 2012 data \cite{CMS-bb}, a signal strength  $\mu_{bb} \approx 1.5$
has been found in the $Z\to \nu \bar \nu$ and $Z \to \ell^+\ell^-$ channels with
a large error bar. Very roughly, one can assume that the additional events from
the $A\to Zh$ channel should be observed if they exceed this sensitivity when
extrapolated to include the full 2012 data.   

One should note that the information from the $pp \to Zh$  search in the SM
provides  only a lower limit for the sensitivity as in the present case one can
benefit from the fact that  the invariant mass of the four fermion final
state (without neutrinos)  which should peak at the value $M_A$ will  further
suppress the continuum background, in particular the $Z+b \bar b$ events.
However, as $h$ is originating from the decay of the state $A$  which should not
be very heavy,  it  has not enough transverse momentum to
strengthen  the boosted jet techniques that allow to isolate the $h \to b\bar
b$  signal from the QCD background.

\subsubsection{The $\mathbf{H \to hh}$ channel}
  
The channel $pp \to H\to hh$ is similar to  $A\to hZ$:  it  has very
large production rates in the low $\tb$ regime in the mass range  $250~{\rm
GeV}  \lsim M_H \lsim 350~{\rm GeV}$ when the decay channels $H\to hh$ is
kinematically open and the $H\to t \bar t$ mode is closed; the $gg \to H$ cross
section should be substantial in this area of the parameter space.  

If the dominant $h\to b\bar b$ decay is considered, the signal topology has some
similarities with that of the process $gg \to b\bar b+A/H$ which was discussed
here as being one of the main MSSM Higgs processes at high $\tb$ and  searched
for by the CMS collaboration with  the 7 TeV data \cite{CMS-bbbb}.  However, the
kinematical behavior is very different and in the signal events, one can use 
further constraints,  $M_{bb} \approx M_h$ and $M_{bbbb} \approx M_H$ (see
Ref.~\cite{ABM} where a characterization of this channel has been made). In
fact, the $H\to hh$  channel has more similarities with double production of the
SM--like Higgs boson, $gg \to hh$, which is considered for the measurement of 
the Higgs self--coupling a  the 14 TeV LHC with a high luminosity. This process
has been revisited recently \cite{hhh} and it has been shown that the final
state  channels $b\bar b \tau \tau$ and $b\bar b\gamma\gamma$ would be viable at
$\sqrt s=14$ TeV and ${\cal L} \gsim 300$  fb$^{-1}$. Because the $h \to
\gamma\gamma$ decay is too rare,  only the first  process could be considered at
$\sqrt s=8$ TeV with 25 fb$^{-1}$ data. Note that here again, one could use the
reconstructed $H$ mass constraint, $M_{H}=M_{hh}$,  to further suppress the
continuum background.

\subsection{Expectations for the LHC at 8 TeV}

It is obvious that a truly reliable estimate of the sensitivity on the heavy
neutral MSSM Higgs bosons in the various channels discussed before can only come
from the  ATLAS and CMS collaborations. We will nevertheless attempt in this
section to provide a very rough estimate of the achievable sensitivities  using
present searches conducted for a heavy SM Higgs and in beyond the SM scenarios.
The very interesting results that could be obtained would hopefully convince the
experimental  collaborations to conduct analyses in this area. 

Following the previous discussions, our working assumptions to derive the
possible sensitivities in the various considered search channels are as
follows: 

-- $H\to WW,ZZ$: we will use the recently published CMS analysis of
Ref.~\cite{CMS-VV} that has been performed with the $\approx 10$ fb$^{-1}$  data
collected in the 7+8 TeV runs and in which all possible channels $H\to ZZ \to
4\ell, 2\ell 2\nu, 2\ell jj, 2\nu jj$ and $H\to WW \to 2\ell 2\nu, \ell \nu jj$
have been included and combined. In the entire range $M_H=160$--350 GeV,  where the SM Higgs
boson almost exclusively  decays into $WW$ or $ZZ$ states, we will assume the
cross section times decay branching ratio upper limit that has been given in 
this CMS study,

-- $H/A \to t\bar t$: we make use of  the ATLAS \cite{ATLAS:1207.2409} and CMS
\cite{CMS:1211.3338} searches at $\sqrt s=7$ TeV for new $Z'$ or Kaluza--Klein
gauge bosons that decay into $t\bar t$ pairs in the  lepton+jets final state
topology. Considering  a small total width for the resonance, limits on the
cross sections times branching ratio of $\approx 6, 3$ and $0.75$ pb  for a
resonance mass of, respectively, 350, 500 and 1000 GeV are assumed. This is
equivalent to a sensitivity that varies with $1/M_{tt}^2$ that we will 
optimistically assume to also cover the low mass resonance range. 

-- $A \to hZ$: we will use the sensitivity given by ATLAS \cite{ATLAS-bb} and
CMS \cite{CMS-bb} in their search for the SM Higgs--like strahlung process
$pp\to hZ$ with $h\to b\bar b$ and $Z\to \ell \ell, \nu\bar \nu$, $\sigma
/\sigma^{\rm SM}=2.8$ with 17 fb$^{-1}$ data at $\sqrt s=7+8$ TeV (we have
included the error bar). This should be sufficient as, in addition, we would
have  on top the constraint    from the reconstructed mass in the $\ell \ell
b\bar b$ channel which is not used in our analysis. 

-- $H\to hh$: we use the analysis of the process $gg \to hh$ in the SM performed
in Ref.~\cite{hhh} for the 14 TeV LHC that we also scale down to the current
energy and luminosity. The final state $bb \tau\tau$ final state will be
considered, with the assumption that the cross section times branching ratio
should be larger than $\sigma \times {\rm BR} \sim 50$ fb for
illustration.    

\begin{figure}[!h]
\begin{center}
\vspace*{-5mm}
\epsfig{file=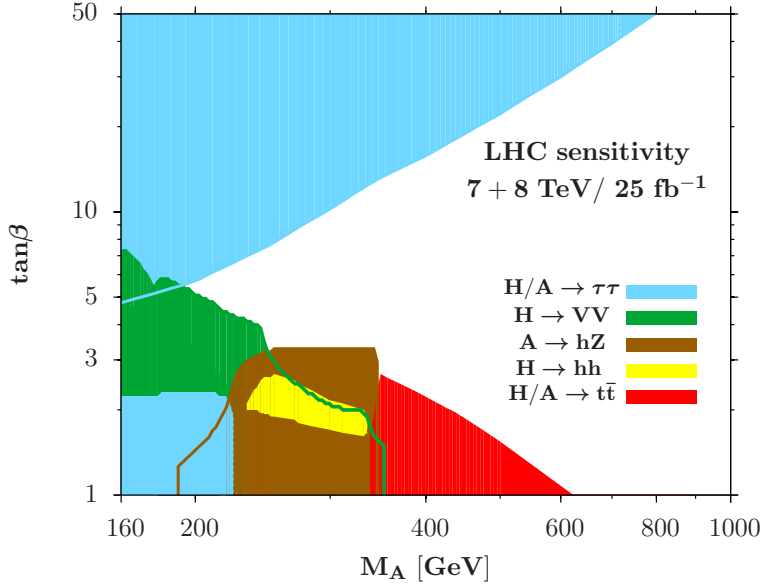,height=7.7cm} 
\end{center}
\vspace*{-5mm}
\caption[]{The estimated sensitivities in the various search channels for the 
heavier MSSM Higgs bosons in the $[\tb,M_A]$ plane: $H/A \to \tau^+ \tau^-$
(light blue), $H \to WW+ZZ$ (green), $H/A
\to t\bar t$ (red), $A\to hZ$  (brown) and $H \to hh$ (yellow). The projection
is made for the LHC with 7+8 TeV and the full 25 fb$^{-1}$ of data 
collected so far. The radiative corrections are such that the lightest
$h$ mass is $M_h=126$ GeV.}
\label{Fig:sensitivity}
\vspace*{-2mm}
\end{figure}

The results are shown in Fig.~\ref{Fig:sensitivity} with an extrapolation to the
full 25 fb$^{-1}$ data of the 7+8 TeV LHC run. Again, we assumed  that the
sensitivity scales simply as the  square root of the number of events. The
sensitivities from the usual $H/A \to \tau^+\tau^-$ channel is also shown. 
The green and red areas correspond to the domains
where the $H\to VV$ and $H/A \to t\bar t$ channels become   constraining with
the assumptions above. The sensitivities in the $H\to hh$ and $A\to hZ$ modes
are  given by, respectively, the yellow and brown areas that peak in the mass
range $MA=250$--350 GeV visible at very low $\tb$ values. We refrain from
extrapolating to the LHC with 14 TeV c.m. energy. 

The outcome is impressive. These channels, in particular the $H \to VV$ and $H/A
\to t \bar t$ processes, are very constraining as they cover the entire low
$\tb$ area that was previously excluded by the LEP2 bound up to $M_A \approx
500$ GeV. Even  $A \to hZ$ and $H \to hh$ are visible in small portions of the
parameter space at the upgraded LHC.

\subsection{Remarks on the charged Higgs boson}

We close this discussions with a few remarks on the charged Higgs boson case. 
First of all, the production rates are very large only for $M_{H^\pm} \lsim 170$
GeV when the $H^\pm $ state can be produced in top decays. In this case, the
decay channel $H^\pm \to \tau \nu$ is always substantial and leads to the
constraints that have been discussed earlier and which are less effective than 
those coming from $H/A \to \tau \tau$ searches at high $\tb$. In the low $\tb$
region, two other channels can be considered: $H^+ \to c\bar s$ that has been
studied by the ATLAS collaboration in a two--Higgs doublet model with the 7 TeV
data  \cite{LHC-cs} and $H^+ \to c\bar b$. The branching ratio for the latter
channel is significant for $\tb \lsim 3$ and has been obtained by assuming the
same CKM angles as in the SM, in particular $V_{cb} \approx 0.04$ \cite{PDG}.
This channel, if observed would thus allow to check some of the CKM  matrix
elements in the charged Higgs sector. 

\begin{figure}[!h]
\begin{center}
\vspace*{-5mm}
\epsfig{file=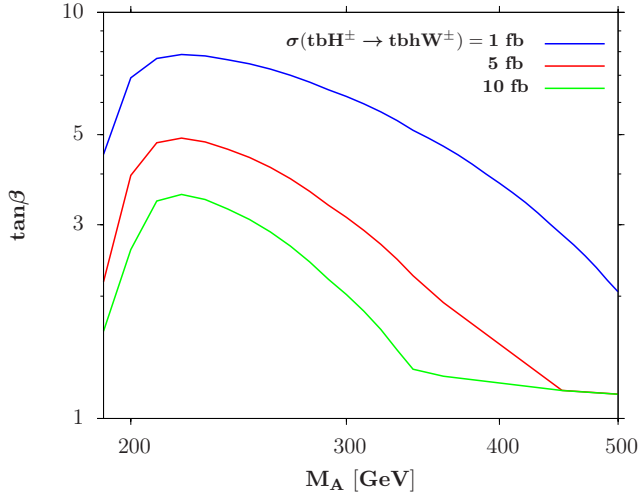,height=6.5cm} 
\end{center}
\vspace*{-5mm}
\caption[]{The production cross sections times decay branching ratio 
at the LHC with $\sqrt s=14$ TeV for the process $pp \to t\bar bH^-+ \bar t 
bH^+$  with $H^\pm \to hW^\pm$ in the $[\tb,M_A]$ plane. The contours are
for some (rough estimates of) limiting values of $\sigma \times {\rm BR}$.}
\label{Fig:H+}
\vspace*{-5mm}
\end{figure}

 Finally, the processes $t \to H^+b$ at low mass and $pp \to bt H^\pm$ at high
mass with $H^\pm \to W h$ can have large rates at sufficiently low $\tb$.  The
cross section times branching fraction is displayed in Fig.~\ref{Fig:H+}  in the
$[\tb,M_A]$ plane for a 14 TeV c.m. energy.  Shown are the contours with 
$\sigma \times {\rm BR}=1, 5$ and 10 fb which, for a luminosity of 300
fb$^{-1}$ would correspond to a small number of events. We will not perform  an
analysis for this particular final state. We simply note that the final state
topology, $pp \to tb H^\pm \to tb Wh $ resembles that of the $pp \to t\bar t h$
process that is considered as a means to measure the $ht\bar t$ Yukawa coupling
and which  is considered to be viable at 14 TeV with a high luminosity.

Hence, even for the charged Higgs bosons, there are interesting search channels
which can be considered if the low $\tb$ region is reopened.

\newpage

\section{Conclusions}

After the observation of the 126 GeV SM--like Higgs boson by the ATLAS and CMS
collaborations, the next challenge at the LHC should be to search for new 
phenomena beyond the SM. This can be done not only by refining the precision
determination of the properties of the observed Higgs particle to pin down small
deviations of its couplings from the SM expectations, but also by looking for
the direct production of new states. 

In this paper, we have considered the production of the heavier $H,A$ and
$H^\pm$ bosons of the MSSM at the LHC, focusing on the low $\tb$ regime,    $\tb
\lsim $3--5.   We have first shown that this area of the MSSM parameter space,
which was long thought  to be excluded, is still viable provided that the SUSY
scale is assumed  to be very high, $M_S \gsim 10$ TeV. For such $M_S$ values,
the usual tools that allow to determine the masses and couplings of the Higgs
and SUSY particles in the MSSM, including the higher order corrections,  become
inadequate.  We have used a simple but not too inaccurate  approximation to
describe the radiative corrections to the Higgs sector, in which the unknown
scale $M_S$ and stop mixing parameter $X_t$ are traded against the measured $h$
boson mass, $M_h \approx 126$ GeV. One would  then have, to a good
approximation, only two basic input parameters in the MSSM Higgs sector even at
higher orders: $\tb$ and $M_A$,  which can take small values, $\tb \approx 1$
and $M_A = {\cal O}(200)$ GeV,  provided that $M_S$ is chosen to be sufficiently
large. 

In the low $\tb$ region, there is a plethora of new search channels for the
heavy MSSM Higgs bosons that can be analyzed at the LHC. The neutral  $H/A$
states  can be still be produced in the gluon fusion mechanism with large rates,
and they will decay  into a variety of interesting final states such as  $H\to
WW,ZZ$, $H \to hh$, $H/A  \to t \bar t$, $A\to hZ$. Interesting decays can also
occur in  the case of the charged Higgs bosons, e.g.    $H^+ \to hW, c\bar s,
c\bar b$. These modes come in addition to the two channels $H/A \to 
\tau^+\tau^-$ and $t \to bH^+ \to b\tau \nu$ which are currently being studied
by ATLAS and CMS and which are very powerful  in constraining the parameter 
space at high $\tb$ values and, as is shown here, also at low $\tb$ values.

We have shown that already with the current LHC data at $\sqrt s=7$+8 TeV, the
area with small $\tb$ and $M_A$ values can be probed by simply  extrapolating to
the MSSM Higgs sector the  available analyses in the search of the SM Higgs
boson at high masses in the $WW$ and $ZZ$ channels and the limits obtained  in
the $t\bar t$ channels in  the search for high--mass new gauge bosons from
extended gauge or extra--dimensional theories. The sensitivity in these channels
will be significantly enhanced at the 14 TeV LHC run once 300 fb$^{-1}$ data
will be collected.  In the absence of any signal at this energy, the $[\tb,
M_A]$ plane can be entirely closed for any $\tb$ value and a pseudoscalar mass
below $M_A \approx 500$ GeV.  Additional and complementary searches can also be
done in the charged Higgs case in channels that have not been studied so far
such as $H^+ \to Wh$ but  we did not analyze this issue in detail. 

Hence, all channels that have been considered for the SM Higgs boson in the high
mass range, plus some processes that have been considered  for other new physics
searches,  can be recycled for the search of the heavier MSSM Higgs bosons in
the low $\tb$ regime. For instance, many of these MSSM Higgs processes could
benefit from the current  searches of multi--lepton events with missing energy
in SUSY theories. As in all channels we have $W,Z$ and additional $h$ bosons in
the  final states, multileptons and missing energy are present in most of the 
topologies. One could then use the direct  searches for SUSY particles such as
charginos and neutralinos to probe also the  MSSM heavier Higgs states. 

All this promises a very nice and exciting program for Higgs searches at the LHC
in both the present and future runs. One could then cover the entire MSSM
parameter space: from above (at high $\tb$) by improving  the $H/A \to \tau\tau$
searches but also from below (at low $\tb$) by using the $WW,ZZ, tt, ..$
searches. The coverage of the  $[\tb, M_A]$ plane will be done in a model
independent way, with no assumption on $M_S$ and possibly on  any other SUSY
parameter\footnote{This approach is orthogonal to that of 
Ref.~\cite{benchmark-update} in which specific benchmark scenarios with fixed
SUSY parameters (which might need to be updated soon) are proposed.  We note
that for all the proposed benchmarks scenarios \cite{benchmark-update}, the SUSY
scale  is fixed to $M_S= 1$ or 1.5 TeV which excludes the low (and possibly
intermediate) $\tb$ regime and, hence, the possibility of discussing the
processes analysed here.}. The indirect information from the lighter Higgs mass 
will be included as well as the information from the Higgs couplings, as the
sensitivity regions cover also that which are excluded from the measurement of
the $h$ properties at the LHC.  

One can of course use these channels in other extensions of the SM. An example
would be SUSY extensions beyond the MSSM where $M_h$ can be made large enough
without  having large $M_S$ values; this  is the case of the NMSSM where the
maximal $M_h$ value can be obtained at $\tb \approx 2$ \cite{NMSSM}. Another
example would be a non--SUSY two--Higgs doublet model  where there is more
freedom in the parameters space and all channels analyzed here and even some
more could be  relevant; discussions along these lines have already started
\cite{ATLAS-2HDM,2HDM}).  The numerous search channels discussed in this
paper  might allow to probe in a more comprehensive manner the extended
parameter space of these models.  \bigskip

\noindent {\bf Acknowledgements}:  
AD thanks the CERN theory division for the kind hospitality offered to him. 
This work is supported by the ERC Advanced Grant Higgs@LHC.


\newpage

\begin{thebibliography}{999} 

\bibitem{Discovery} The ATLAS collaboration, Phys. Lett. B716 (2012) 1;  the 
CMS collaboration, Phys. Lett. B716 (2012) 30. 

\bibitem{Higgs} P. Higgs, Phys. Lett. 12 (1964) 132; F.  Englert and  R. Brout,
Phys. Rev. Lett. 13 (1964) 321; G. Guralnik, C. Hagen and T. Kibble, Phys. Rev.
Lett. 13 (1964) 585; P. Higgs. Phys. Rev. 145 (1966) 1156. 

\bibitem{HHG} J. Gunion, H. Haber, G. Kane and S. Dawson, ``The Higgs Hunter's
Guide", Reading 1990.

\bibitem{Review1} A.~Djouadi,  Phys. Rept. 457 (2008) 1, [arXiv:hep-ph/0503172].


\bibitem{SUSY}  H. Haber and G. Kane, Phys. Rep. 117 (1985) 75; S. Martin,
hep-ph/9709356;   M. Drees, R. Godbole and P. Roy,  {\it Theory and
phenomenology of sparticles}, World Scientific, 2005. 


\bibitem{pMSSM} A. Djouadi et al. (MSSM Working Group), hep-ph/9901246.

\bibitem{Review2} A.~Djouadi,  Phys. Rept. 459 (2008) 1, [arXiv:hep-ph/0503173].

\bibitem{CR-review} M. Carena and H.  Haber, Prog. Part. Nucl.  Phys. 50 (2003)
63; S.~Heinemeyer, W.~Hollik and G.~Weiglein, Phys. Rept. 425 (2006) 265; B.C.
Allanach et al., JHEP 0409 (2004) 044. 

\bibitem{paper1} A. Arbey et al., Phys. Lett. B708 (2012) 162.

\bibitem{LHC-all} See the talks given by the members of the ATLAS and CMS
collaborations 
at the XLVIIIth ``Rencontres de Moriond'',  2-16 March 2013, La Thuile, Italy.

\bibitem{decoupling} See for instance, H.E. Haber, CERN-TH/95-109 and
hep-ph/9505240.

\bibitem{fine-tuning} See e.g. R. Barbieri and G.F. Giudice, Nucl. Phys. B306
(1988) 63.

\bibitem{LEP2} The LEP collaborations, Eur.Phys.J. C47 (2006) 547.

\bibitem{split}  N. Arkani-Hamed and S. Dimopoulos, JHEP 0506 (2005) 073; G.F.
Giudice and A. Romanino, Nucl. Phys. B699 (2004) 65; J.D. Wells, Phys. Rev. D71
(2005) 015013.

\bibitem{high-scale} See e.g. L.J. Hall and Y. Nomura, JHEP 1003 (2010) 076. 

\bibitem{NUHM}  J. R. Ellis et al., Nucl. Phys. B 652 (2003) 259;  H. Baer et
al., Phys. Rev. D 71 (2005) 095008; J. R. Ellis, K. A. Olive and P. Sandick,
Phys. Rev. D 78 (2008) 075012; L. Roszkowski et al., Phys. Rev. D 83 (2011)
015014. For a recent account, see e.g. S.S. AbdusSalam et al., Eur. Phys. J. C71
(2011)  1835. 

\bibitem{slim} A. Delgado and G.F. Giudice, Phys. Lett. B627 (2005) 155;
	E. Arganda, J.L. Diaz-Cruz and A. Szynkman, arXiv:1211.0163 and 
	arXiv:1301.0708.


\bibitem{Suspect} A.~Djouadi, J.L.~Kneur and G.~Moultaka, Comput. Phys. Commun.
176 (2007)  426.

\bibitem{RGE}  H.~Baer, F.~Paige, S..~Protopopescu and X.~Tata,
hep-ph/0001086; B.~Allanach,  Comput.\ Phys.\ Commun. 143 (2002) 305; 
W.~Porod, Comput.\ Phys.\ Commun. 153 (2003) 275. 

\bibitem{bds} N. Bernal, A. Djouadi and P. Slavich, JHEP 0707 (2007) 016.

\bibitem{prepa} E. Bagnaschi, N. Bernal, A. Djouadi, J. Quevillon and P.
Slavich, in preparation.  



\bibitem{CR-1loop} Y. Okada, M. Yamaguchi and T. Yanagida, Prog. Theor. Phys. 
85 (1991) 1;  J. Ellis, G. Ridolfi and F.~Zwirner, Phys. Lett. B257 (1991) 83;
H.E.~Haber and R.~Hempfling, Phys. Rev. Lett. 66  (1991) 1815.

\bibitem{Maiani:2012ij}
L.~Maiani, A.~Polosa and V.~Riquer, New J.\ Phys.\   14 (2012) 073029; 
Phys.\ Lett.\ B718 (2012) 465.

\bibitem{tb-review} See, for instance, F. Zwirner, hep-ph/9203204. 

\bibitem{PBMZ} D.~Pierce, J.~Bagger, K.~Matchev and R.~Zhang, Nucl. Phys. B491
(1997) 3.

\bibitem{CR-eff} M.~Carena, J.R.~Espinosa, M.~Quiros and C.E.~Wagner, Phys. 
Lett. B355 (1995) 209; H.~Haber, R.~Hempfling and A.~Hoang, Z. Phys. C75 (1997)
539.

\bibitem{CR-2loop} G. Degrassi, P. Slavich, F. Zwirner, Nucl. Phys. B611 (2001)
403; A. Brignole, G. Degrassi, P. Slavich and F. Zwirner, Nucl. Phys. B631 (2002)
195; Nucl. Phys. B643 (2002) 79. 

\bibitem{CR-FH} S.~Heinemeyer, W.~Hollik and G.~Weiglein, Phys. Rev. D58 (1998)
091701; Eur.\ Phys.\ J.C9 (1999) 343. 

\bibitem{CR-3loop} P. Kant, R.V. Harlander, L. Mihaila and M. Steinhauser, JHEP
 1008 (2010) 104.	

\bibitem{paper2} A. Arbey, M. Battaglia, A. Djouadi and F. Mahmoudi, 
JHEP 1209 (2012) 107.

\bibitem{paper3} A. Arbey, M. Battaglia, A. Djouadi and F. Mahmoudi,  
Phys. Lett. B720 (2013) 153.

\bibitem{benchmark} M. Carena, S. Heinemeyer, C. Wagner and G. Weiglein, Eur.
Phys. J. C26 (2003) 601. 

\bibitem{benchmark-update} M. Carena et al., arXiv:1302.7033. 

\bibitem{Feynhiggs} S.~Heinemeyer, W.~Hollik and G.~Weiglein, Comp. Phys.
Commun. 124 (2000) 76.
	
\bibitem{PDG} J. Beringer (Particle Data Group) et al., Phys. Rev. D86 (2012)
010001.

\bibitem{alekhin-fate} S. Alekhin, A. Djouadi and S. Moch,  Phys. Lett. B716
(2012) 214.

\bibitem{strumia} G. Giudice and A. Strumia, Nucl.Phys. B858 (2012) 63.
		
\bibitem{focus} J. Feng, K. Matchev and T. Moroi,  Phys. Rev. Lett. 84 (2000)
2322; Phys. Rev. D61 (2000) 075005; J.L. Feng, arXiv:1302.6587 [hep-ph]. 

\bibitem{mtheory} See eg.g. D. Feldman, G. Kane, E. Kuflik and R. Lu, 
Phys. Lett. B704 (2011) 56; S. Akula, B. Altunkaynak, D. Feldman,   Pran Nath and
G. Peim, Phys. Rev. D85 (2012) 075001 B. Acharya, G. Kane and P. Kumar, Int. J.
Mod. Phys. A27 (2012) 1230012; G. Kane, R. Lu and B. Zheng, Int. J. Mod.Phys.
A28 (2013) 1330002.

\bibitem{CR-deltab} See e.g., M.~Carena, D.~Garcia, U.~Nierste and C.E.~Wagner,
Nucl. Phys. B577 (2000) 88; D. Noth and M. Spira, Phys. Rev. Lett. 101 (2008)
181801. 

\bibitem{DKMZ} A. Djouadi, W. Kilian, M.M. Muhlleitner and P.M. Zerwas,  Eur.
Phys. J. C10 (1999) 27; Eur. Phys. J.   C10 (1999) 45.

\bibitem{intense} E. Boos et al., Phys. Rev. D66 (2002) 055004;    E. Boos, A.
Djouadi and A. Nikitenko, Phys. Lett. B578 (2004) 384. 

\bibitem{Baglio2} J. Baglio and A. Djouadi, JHEP 1103 (2011) 055. 

\bibitem{LHCWG} S. Dittmaier et al. (LHC Higgs Working Group),
arXiv:1101.0593 [hep-ph].

\bibitem{scott} D. Dicus and S. Willenbrock, Phys. Rev. D39 (1989) 751. 

\bibitem{ggH-NLO} M.  Spira, A. Djouadi, D. Graudenz and P.M. Zerwas,  Nucl. 
Phys.  B453 (1995) 17.

\bibitem{HIGLU} M. Spira, Fortschr.~Phys.~46 (1998) 203; hep-ph/9510347.

\bibitem{Michael-web} Michael Spira site: {\tt http://mspira.home.cern.ch/
mspira/proglist.html}.


\bibitem{bbH-NLO} J. Campbell, R. K. Ellis, F. Maltoni and S.  Willenbrock,
Phys. Rev. D67 (2003) 095002;  F. Maltoni, Z. Sullivan and S. Willenbrock Phys.
Rev.  D67 (2003) 093005.
	
\bibitem{bbH-NNLO} R. Harlander and W. Kilgore, Phys.  Rev. D68 (2003) 013001.

\bibitem{SUSHI} R. Harlander, S. Liebler and H. Mantler, Comp. Phys.
Comm. 184 (2013) 1605.

\bibitem{ggbbH-NLO}S. Dittmaier, M. Kr\"amer and M. Spira, Phys. Rev. D70 (2004)
074010;  S. Dawson et al,  Phys. Rev. D69 (2004) 074027.


\bibitem{ABM} 	A. Arbey, M. Battaglia and F. Mahmoudi, arXiv:1303.7450.


\bibitem{ggH-squarks} S. Dawson, A. Djouadi and M. Spira, Phys. Rev. Lett. 77
(1996) 16; R. Harlander and M. Steinhauser, JHEP 0409 (2004) 066, ibid. Phys.
Rev. D68  (2003) 111701; M. Muhlleitner, H. Rzehak and M. Spira,  JHEP 0904
(2009) 023.

\bibitem{Liu:2012qu}
N.~Liu, L.~Wu, P.~W.~Wu and J.~M.~Yang, 
JHEP 1301 (2013) 161.
  
\bibitem{LHC-SUSY} ATLAS Collaboration, ATLAS-CONF-2012-105 and
ATLAS-CONF-2012-108;  CMS Collaboration, arXiv:1301.0916 and  arXiv:1212.6428. 


\bibitem{LHC-mu} ATLAS Collaboration, ATLAS-CONF-2013-010 and 
arXiv:1211.6956; CMS Collaboration, CMS-HIG-12-011.

\bibitem{CMS-bbbb} CMS Collaboration, CMS-PAS-HIG-12-033.


\bibitem{MSTW}  A.D. Martin, W. Strirling, R. Thorne and G. Watt,  Eur. Phys. J.
C63 (2009) 189.


\bibitem{ggH-NLO-approx} A. Djouadi, M. Spira and P.M. Zerwas, Phys. Lett. B264
(1991) 440; S.  Dawson, Nucl. Phys. B359 (1991) 283,  M. Spira et al.,   Phys.
Lett.  B318 (1993) 347.

\bibitem{ggH-NNLO}  R.V.~Harlander and W. Kilgore, Phys. Rev. Lett. 88 (2002)
201801; C. Anastasiou and K. Melnikov, Nucl. Phys. B646 (2002) 220; V.
Ravindran, J. Smith and W.L. Van Neerven, Nucl. Phys. B665 (2003) 325; R.
Harlander and W. Kilgore, JHEP 0210 (2002) 017. S.~Catani et al., JHEP 0307
(2003) 028.

\bibitem{LO-H+} A. Bawa, C. Kim and A. Martin, Z. Phys. C47 (1990) 75; V. Barger,
R. Phillips and D.P. Roy, Phys. Lett. B324 (1994) 236;  S. Moretti and K.
Odagiri, Phys. Rev. D55 (1997) 5627; J. Gunion, Phys. Lett. B322 (1994) 125;  F.
Borzumati, J.L. Kneur and N. Polonsky, Phys. Rev. D60 (1999) 115011. 


\bibitem{NLO-H+} T. Plehn, Phys. Rev. D67 (2003) 014018.

\bibitem{Stefano} For a review, see: S. Moretti, Pramana 60 (2003) 369.

\bibitem{HDECAY} A.~Djouadi, J.~Kalinowski and M.~Spira, Comput. Phys. Commun.
108 (1998) 56. An update of the program with M. Muhlleitner in addition
appeared  in hep-ph/0609292.

\bibitem{H3body} A. Djouadi, J. Kalinowski and P. Zerwas, Z. Phys. C70
(1996) 435; S. Moretti,  J. Stirling, Phys. Lett. B347 (1995) 291;
F. Borzumati and A. Djouadi, Phys. Lett. B549 (2002) 170.


\bibitem{ATLAS-gg} The ATLAS collaboration, ATLAS-CONF-2013-014.

\bibitem{CMS-gg} The CMS collaboration,	CMS-PAS-HIG-13-001

\bibitem{ATLAS-ZZ} The ATLAS collaboration, ATLAS-CONF-2013-013. 

\bibitem{CMS-ZZ} The CMS collaboration, CMS-PAS-HIG-13-002.

\bibitem{ATLAS-WW} The ATLAS collaboration, ATLAS-CONF-2013-030.  	 

\bibitem{CMS-WW} The CMS collaboration, CMS-PAS-HIG-13-003.
 	 
\bibitem{ATLAS-bb} The ATLAS collaboration,	 ATLAS-CONF-2012-161.

\bibitem{CMS-bb} The CMS collaboration, CMS-PAS-HIG-12-044.

\bibitem{ATLAS-tau} The ATLAS collaboration, 	 ATLAS-CONF-2012-094 and arXiv:1211.6956.

\bibitem{CMS-tau}  The CMS collaboration, CMS-PAS-HIG-12-050.

\bibitem{ATLAS-H+}  The ATLAS collaboration, ATLAS-CONF-2012-011 and arXiv:1204.2760.

\bibitem{CMS-H+} The CMS collaboration, CMS-HIG-11-019 and  arXiv:1205.5736.

\bibitem{ratios}   A. Djouadi, arXiv:1208.3436; A. Djouadi and G. Moreau,
arXiv:1303.6591. 

\bibitem{talks} A. Djouadi, see talks given in e.g. ``Rencontres de la Vall\'ee
d'Aoste", La Thuile, February 2013; ``Discrete 2012", Lisbon, December 2012; 
CMS Exotic Higgs group meeting, CERN, February 2013; ATLAS MSSM Higgs group
meeting, CERN, December 2012.    

\bibitem{others} See also M. Carena, P. Draper, T. Liu and  C. Wagner, Phys. Rev.  D84 (2011)
095010;  N.  Christensen, T. Han and  S. Su, Phys. Rev. D85 (2012) 115018; J.
Chang, K. Cheung, P. Tseng and T-C. Yuan, Phys.Rev. D87 (2013)  035008.

\bibitem{CMS-VV} The CMS collaboration, arXiv:1304.0213.

\bibitem{ATLAS-2HDM} The ATLAS collaboration, ATLAS-CONF-2013-027.  	 

\bibitem{ATLAS:1207.2409} The ATLAS collaboration, arXiv:1207.2409.

\bibitem{CMS:1211.3338} The CMS collaboration, arXiv:1211.3338.

\bibitem{ttH-boost} T. Plehn, G.P. Salam and M. Spannowsky, Phys. Rev. Lett. 104
(2010) 111801.

\bibitem{VH-boost} J. Butterworth, A. Davison, M. Rubin and G. Salam, Phys. Rev.
 Lett. 100 (2008)  242001.
 
\bibitem{hhh} J.~Baglio, A.~Djouadi, R.~Gr\"ober, M.~M.~M\"uhlleitner, J.~Quevillon and M.~Spira,
JHEP 1304 (2013) 151. See also : 
M.~J.~Dolan, C.~Englert and M.~Spannowsky,
 JHEP 1210 (2012) 112.
  
\bibitem{LHC-cs} The ATLAS collaboration, ATLAS-CONF-2011-094 and
arXiv:1302.3694.

\bibitem{NMSSM} U. Ellwanger, C. Hugonie and  A.M. Teixeira, Phys. Rept. 496
(2010) 1; M. Maniatis, Int.J.Mod.Phys. A25 (2010) 3505; A.~Djouadi  et al., JHEP
0807 (2008) 002. For recent discussions, see: G. B\'elanger et al.,  JHEP 1301
(2013) 069; S. King, M. Muhlleitner and R. Nevzorov, Nucl.Phys. B860 (2012)
207.

\bibitem{2HDM}  G. Branco et al.,  Phys. Rept. 516 (2012) 1.  For recent
discussions, see:   
W.~Altmannshofer, S.~Gori and G.~D.~Kribs,
Phys.\ Rev.\ D  86 (2012) 115009;
Y. Bai et al., arXiv:1210.4922;  A. Drozd et al., arXiv:
1211.3580; J. Chang et al., arXiv:1211.3849; B. Grinstein and P. Uttayarat,
arXiv:1304.0028.
  
\end{thebibliography}
\end{document}